\documentclass[twocolumn,pra,showpacs]{revtex4}
\usepackage{amssymb}
\usepackage{amsfonts}
\usepackage{amsmath}
\usepackage{graphicx}

\begin{document}

\title{Post-Markovian quantum master equations from classical environment
fluctuations}
\author{Adri\'{a}n A. Budini}
\affiliation{Consejo Nacional de Investigaciones Cient\'{\i}ficas y T\'{e}cnicas
(CONICET), Centro At\'{o}mico Bariloche, Avenida E. Bustillo Km 9.5, (8400)
Bariloche, Argentina, and Universidad Tecnol\'{o}gica Nacional (UTN-FRBA),
Fanny Newbery 111, (8400) Bariloche, Argentina}
\date{\today }

\begin{abstract}
In this paper we demonstrate that two commonly used phenomenological
post-Markovian quantum master equations can be derived without using any
perturbative approximation. A system coupled to an environment characterized
by self-classical configurational fluctuations, the latter obeying a
Markovian dynamics, defines the underlying physical model. Both
Shabani-Lidar equation [A. Shabani and D. A. Lidar, Phys. Rev. A \textbf{71}%
, 020101(R) (2005)] and its associated approximated integro-differential
kernel master equation are obtained by tracing out two different bipartite
Markovian Lindblad dynamics where the environment fluctuations are taken
into account by an ancilla system. Furthermore, conditions under which the
non-Markovian system dynamics can be unravelled in terms of an ensemble of
measurement trajectories are found. In addition, a non-Markovian quantum
jump approach is formulated. Contrary to recent analysis [L. Mazzola, E. M.
Laine, H. P. Breuer, S. Maniscalco, and J. Piilo, Phys. Rev. A \textbf{81},
062120 (2010)], we also demonstrate that these master equations, even with
exponential memory functions, may lead to non-Markovian effects such as an
environment-to-system backflow of information if the Hamiltonian system does
not commutate with the dissipative dynamics.

\end{abstract}

\maketitle



\section{Introduction}

Contrary to Markovian Lindblad dynamics \cite{alicki,breuerbook}, the
description of non-Markovian open quantum systems relies on density matrix
evolutions defined by integro-differential equations \cite{haake}. In these
dynamics, the dependence of the system state on its previous history is
weighted by a memory kernel function, which in turn may itself depend on
each dissipative channel. Different theoretical approaches and physical
situations had been analyzed by many authors in order to establish and
characterize these equations \cite%
{stenholm,classBu,giraldi,VacCol,tanos,wilkie,salo,cresser,Lidar,ManisPetru,unMedio,piilo,grano,LindbladRate,vacchini,SemiMarkov,Kosa,NoMeasure}%
. The unravelling of the non-Markovian dynamics in terms of measurement
trajectories has also been extensively studied \cite%
{diosi,maniscalco,petruccione,barchielli,BuJumpSMS,OneChannel,Embedding}.

On the basis of a phenomenological measurement theory Shabani and Lidar
introduced a non-Markovian dynamics \cite{Lidar}, called post-Markovian
equation, where a single kernel weights the memory effects. In the
\textquotedblleft stationary case\textquotedblright\ it is%
\begin{equation}
\frac{d}{dt}\rho _{t}^{s}=\mathcal{C}_{s}\int_{0}^{t}dt^{\prime
}k(t-t^{\prime })\exp [(t-t^{\prime })\mathcal{C}_{s}]\rho _{t^{\prime
}}^{s},  \label{LidarEquation}
\end{equation}%
where $\rho _{t}^{s}$ is the system density matrix and $\mathcal{C}_{s}$\ is
an arbitrary (diagonalized) Lindblad superoperator,%
\begin{equation}
\mathcal{C}_{s}[\rho ]=\frac{1}{2}\sum_{\alpha }\gamma _{\alpha }([V_{\alpha
},\rho V_{\alpha }^{\dag }]+[V_{\alpha }\rho ,V_{\alpha }^{\dag }]),\ \ \ \
\ \{\gamma _{\alpha }\}\geq 0.  \label{LindbladDefinition}
\end{equation}%
The rates $\{\gamma _{\alpha }\}$\ \textquotedblleft
measure\textquotedblright\ the weight of each dissipative Lindblad channel
defined by the system operators $\{V_{\alpha }\}.$ As mentioned in \cite%
{Lidar}, under the condition $||\mathcal{C}_{s}||\ll 1/t,$ the previous
equation can be approximated as%
\begin{equation}
\frac{d}{dt}\rho _{t}^{s}=\mathcal{C}_{s}\int_{0}^{t}dt^{\prime
}k(t-t^{\prime })\rho _{t^{\prime }}^{s},  \label{OneKernel}
\end{equation}%
which shows the close relationship that exists between both types of
non-Markovian evolutions, Eqs. (\ref{LidarEquation}) and~(\ref{OneKernel}).

Different analysis of both non-Markovian master equations can be found in
literature \cite{ManisPetru,unMedio,piilo,grano}. In Ref. \cite{ManisPetru},
by comparing the solutions of both equations for a qubit system, conditions
on the limit of applicability of each dynamics were established. In Ref. 
\cite{unMedio}, the completely positive condition \cite{alicki,breuerbook}
of the solution maps was studied by assuming an exponential kernel. In that
case, Eq. (\ref{LidarEquation}) always results in a completely positive
solution map while Eq. (\ref{OneKernel}) does not fulfill this condition, in
general \cite{stenholm,classBu}. In Ref. \cite{piilo} it was found that
neither Eq. (\ref{LidarEquation}) nor Eq. (\ref{OneKernel}) are able to
induce \textquotedblleft genuine\textquotedblright\ non-Markovian effects
such as an environment-to-system backflow of information \cite{NoMeasure}. A
stringent constraint on the usefulness of these equations is established by
this result. In addition, the hazards of using non-Markovian evolutions like
Eq. (\ref{LidarEquation}) in systems containing a partially unitary dynamics
were analyzed in Ref. \cite{grano}.

In spite of the previous analysis \cite{ManisPetru,unMedio,piilo,grano},
consistently with its original formulation \cite{Lidar}, the non-Markovian
evolutions (\ref{LidarEquation}) and (\ref{OneKernel}) rely on
phenomenological ingredients such as the memory kernel $k(t).$ In fact,
microscopic dynamics that lead to a given kernel are generally unknown. On
the other hand, assuming that a continuous-in-time measurement process is
performed over the system, it is not known which kind of stochastic
trajectories \cite{plenio,carmichaelbook,carmichael,barchiellibook,bartano}\
may describe the conditional system dynamics \cite%
{diosi,maniscalco,petruccione,barchielli,BuJumpSMS,OneChannel,Embedding}.
The main goal of this paper is to answer these issues. In addition, we
criticize and generalize some previous results \cite%
{ManisPetru,unMedio,piilo,grano}\ about these non-Markovian dynamics.

Over the basis of single molecule spectroscopy arrangements \cite{SMS}, in
the present study we consider a system coupled to an environment
characterized by (Markovian) classical self-fluctuations, which in turn
modify or modulate the system dissipative dynamics. This situation can be
described with a bipartite Lindblad evolution \cite{LindbladRate}, where an
auxiliary ancilla system takes into account the environment fluctuations 
\cite{SMSBudini}. Under these assumptions, without introducing any
perturbative approximation, by using projector techniques we demonstrate
that Eqs. (\ref{LidarEquation}) and (\ref{OneKernel}) describe the system
dynamics for two alternative kinds of system-environment couplings. On the
other hand, generalizing the results of Ref. \cite{OneChannel}, we show
that, under some particular conditions, a non-Markovian quantum jump
approach can be consistently formulated for both non-Markovian master
equations. In fact, under specific symmetries conditions, non-Markovian
evolutions obtained from a partial trace over a bipartite Markovian dynamics
can be unravelled in terms of a set of measurement trajectories whose
dynamics can be written in the (single) system Hilbert space \cite%
{OneChannel}.

We remark that in the present study the Lindblad superoperator (\ref%
{LindbladDefinition}) that defines both master equations (\ref{LidarEquation}%
) and (\ref{OneKernel}) is not a \textquotedblleft
collisional\textquotedblright\ one \cite{Embedding}, that is,%
\begin{equation}
\mathcal{C}_{s}\neq \sum_{\alpha }V_{\alpha }\rho V_{\alpha }^{\dag }-%
\mathrm{I}_{s},\ \ \ \ \ \ \ \ \ \ \sum_{\alpha }V_{\alpha }^{\dag
}V_{\alpha }=\mathrm{I}_{s},  \label{NoColision}
\end{equation}%
where $\mathrm{I}_{s}$\ is the identity matrix. When $\mathcal{C}%
_{s}=\sum_{\alpha }V_{\alpha }\rho V_{\alpha }^{\dag }-\mathrm{I}_{s},$ the
non-Markovian dynamics (\ref{OneKernel}) can be unravelled in terms of a set
of trajectories where the collisional\ superoperator $\mathcal{E}_{s}[\rho
]=\sum_{\alpha }V_{\alpha }\rho V_{\alpha }^{\dag }$ is applied at random
times over the system state \cite{Embedding}. As a consequence of the
assumption (\ref{NoColision}), the results of Ref. \cite{Embedding} do not
apply in the present context.

The paper is structured as follows. In Sec. II, we present the basic
bipartite Markovian model that describes the system-environment coupling.
Both evolutions (\ref{LidarEquation}) and (\ref{OneKernel}) are obtained by
tracing out the configurational bath-fluctuations. In Sec. III, for each
equation we develop a non-Markovian quantum jump approach that allows to
unravel the dynamics in terms of a set of measurement trajectories.
Conditions under which these results can be formulated are found in the
Appendix. In Sec. IV we study an example that exhibits the main features of
the present approach. Furthermore, the model shows that, even with an
exponential memory kernel function, the analyzed post-Markovian quantum
master equations may lead to the development of genuine non-Markovian
effects. In Sec. V we provide the Conclusions.

\section{Classical environment fluctuations}

We consider a system $S$ interacting with an environment that develops
classical self-fluctuations between a set of \textquotedblleft
configurational bath-states\textquotedblright\ \cite{vanKampen}. Each state
corresponds to different Hilbert subspaces of the reservoir which are able
to induce by themselves a different Markovian system dynamics. Hence, the
transitions between them modulate the system evolution \cite{SMS,SMSBudini}.

The transitions between the reservoir-states are defined by a classical
Pauli master equation \cite{vanKampen}%
\begin{equation}
\frac{d}{dt}p_{t}^{i}=\sum_{j}(\phi _{ij}p_{t}^{j}-\phi _{ji}p_{t}^{i}),
\label{classical}
\end{equation}%
where $p_{t}^{i}$ is the probability that at time $t$ the environment is in
the $i-$state $(i=0,1,\cdots i_{\max })$ while $\phi _{ij}$ are the
transition rates $(\phi _{ij}\geq 0,$ $\phi _{ii}=0).$ The system density
matrix $\rho _{t}^{s}$\ is written as%
\begin{equation}
\rho _{t}^{s}=\sum_{i}\rho _{t}^{i},  \label{SUMA}
\end{equation}%
where each auxiliary state $\rho _{t}^{i}$ corresponds to the system state
\textquotedblleft given\textquotedblright\ that the environment is in the $%
i- $state \cite{LindbladRate}. The weight of each bath-state is encoded as $%
p_{t}^{i}=\mathrm{Tr}_{s}\left[ \rho _{t}^{i}\right] .$ The initial
conditions read $\{\rho _{0}^{i}=p_{0}^{i}\rho _{0}^{s}\}.$ The system
dynamics is completely defined after introducing the time evolution of the
states $\rho _{t}^{i}.$ We consider models where the bath-states modulate
(modify) the system dissipative dynamics. In correspondence with Eqs. (\ref%
{LidarEquation}) and (\ref{OneKernel}) two different cases are considered.

\subsubsection*{First case}

In the first case, the evolution of the auxiliary states\ is given by the
\textquotedblleft Lindblad rate equation\textquotedblright\ \cite%
{LindbladRate}%
\begin{equation}
\frac{d}{dt}\rho _{t}^{i}=\mathcal{L}_{s}\rho _{t}^{i}+(1-\delta _{i0})%
\mathcal{C}_{s}\rho _{t}^{i}+\sum_{j}(\phi _{ij}\rho _{t}^{j}-\phi _{ji}\rho
_{t}^{i}),  \label{primerol}
\end{equation}%
where $\delta _{ij}$\ is the Kronecker delta function. The superoperator $%
\mathcal{L}_{s}$ defines the system unitary evolution, but may also include
arbitrary Lindblad contributions. The action of $\mathcal{L}_{s}$\ is
independent of the environment state. On the other hand, the superoperator $%
\mathcal{C}_{s}$ is given by Eq. (\ref{LindbladDefinition}). Its action is
modulated by the environment states. In fact, in Eq. (\ref{primerol}), its
influence is inhibited when the bath is in the state $i=0.$ In any other
case $(i\neq 0),$ the system dynamics also includes the contribution $%
\mathcal{C}_{s}.$ The classical transitions between these regimes is
introduced by the last term in Eq. (\ref{primerol}), which in turn, from $%
p_{t}^{i}=\mathrm{Tr}_{s}\left[ \rho _{t}^{i}\right] ,$ leads to the
classical master equation (\ref{classical}).

\subsubsection*{Second case}

The second dynamics is complementary to the previous one. The auxiliary
states evolution is%
\begin{equation}
\frac{d}{dt}\rho _{t}^{i}=\mathcal{L}_{s}\rho _{t}^{i}+\delta _{i0}\mathcal{C%
}_{s}\rho _{t}^{i}+\sum_{j}(\phi _{ij}\rho _{t}^{j}-\phi _{ji}\rho _{t}^{i}).
\label{segundol}
\end{equation}%
Therefore, here $\mathcal{C}_{s}$ is able to modify the system dynamics only
when the environment is in the state $i=0.$ In any other case $(i\neq 0),$
it is inhibited. Notice that this \textquotedblleft complementary bath
action\textquotedblright\ is the unique difference with Eq.~(\ref{primerol}).

\subsection{Bipartite representation}

The system state (\ref{SUMA}) and the evolution defined by Eqs. (\ref%
{primerol}) and (\ref{segundol}) completely define the (non-Markovian)
system dynamics. Nevertheless, their analysis is simplified if those
dynamics are embedded in a bipartite Markovian dynamics \cite{LindbladRate},
where an extra ancilla (auxiliary) system $(A)$ takes into account the
environment fluctuations. The joint system-ancilla density matrix is denoted
as $\rho _{t}^{sa}.$ From this object, the system state follows from a
partial trace,%
\begin{equation}
\rho _{t}^{s}=\mathrm{Tr}_{a}\left[ \rho _{t}^{sa}\right] =\sum_{i}\left%
\langle a_{i}\right\vert \rho _{t}^{sa}\left\vert a_{i}\right\rangle ,
\label{RhoS_A}
\end{equation}%
where $\{\left\vert a_{i}\right\rangle \},$ $i=0,1\cdots (\dim \mathcal{H}%
_{a}-1)=i_{\max },$ is a complete normalized basis in the ancilla Hilbert
space $\mathcal{H}_{a}.$ Each state $\left\vert a_{i}\right\rangle $\
corresponds to each configurational bath-state. Hence, the auxiliary\ system
states [Eq. (\ref{SUMA})] read%
\begin{equation}
\rho _{t}^{i}=\left\langle a_{i}\right\vert \rho _{t}^{sa}\left\vert
a_{i}\right\rangle ,  \label{auxiliarRho}
\end{equation}%
while the ancilla populations%
\begin{equation}
p_{t}^{i}=\mathrm{Tr}_{s}\left[ \rho _{t}^{i}\right] =\left\langle
a_{i}\right\vert \mathrm{Tr}_{s}\left[ \rho _{t}^{sa}\right] \left\vert
a_{i}\right\rangle ,  \label{populor}
\end{equation}%
define the probability of each configurational bath-state. From now on we
denote indistinctly the bath-states by their ancilla representation.

The time evolution of $\rho _{t}^{sa}$ is given by a Markovian equation that
recovers the previous Lindblad rate equations. It is written as%
\begin{equation}
\frac{d}{dt}\rho _{t}^{sa}=\mathcal{L}[\rho _{t}^{sa}]=(\mathcal{L}_{s}+%
\mathcal{L}_{a}+\mathcal{C}_{sa})\rho _{t}^{sa}.  \label{LindbladBipartito}
\end{equation}%
The superoperator $\mathcal{L}_{s}$ is the same as before and only acts on
the system Hilbert space. The contribution $\mathcal{L}_{a}$ gives the
ancilla dynamics. It introduces the configurational bath-transitions. In
correspondence with the classical evolution (\ref{classical}), it reads%
\begin{equation}
\mathcal{L}_{a}[\rho ]=\frac{1}{2}\sum_{i,j}\phi _{ij}([A_{ij},\rho
A_{ij}^{\dag }]+[A_{ij}\rho ,A_{ij}^{\dag }]).  \label{LindbladClasico}
\end{equation}%
The operators are $A_{ij}=\mathrm{I}_{s}\otimes \left\vert
a_{i}\right\rangle \left\langle a_{j}\right\vert .$

The contribution $\mathcal{C}_{sa}$ is defined in different ways for each
case. In the first case, it must be taken as%
\begin{equation}
\mathcal{C}_{sa}[\rho ]=\frac{1}{2}\sum\nolimits_{i,\alpha }^{^{\prime
}}\gamma _{\alpha }([T_{\alpha i},\rho T_{\alpha i}^{\dag }]+[T_{\alpha
i}\rho ,T_{\alpha i}^{\dag }]),  \label{LindbladInteraction1}
\end{equation}%
where the Lindblad bipartite operators are%
\begin{equation}
T_{\alpha i}=V_{\alpha }\otimes \left\vert a_{i}\right\rangle \left\langle
a_{i}\right\vert .  \label{TAlfa_i}
\end{equation}%
Both the rates $\{\gamma _{\alpha }\}$ and system operators $\{V_{\alpha }\}$
are the same as in Eq. (\ref{LindbladDefinition}). With $\sum%
\nolimits_{i}^{^{\prime }}$ we denote a sum that runs over the states $%
\left\vert a_{i}\right\rangle ,$ $i=1,\cdots (\dim \mathcal{H}_{a}-1),$
excluding the state $\left\vert a_{\mathrm{0}}\right\rangle .$ By using the
definition of the auxiliary states, Eq. (\ref{auxiliarRho}), from the
bipartite dynamics (\ref{LindbladBipartito}), jointly with the definitions (%
\ref{LindbladClasico}) and (\ref{LindbladInteraction1}), it is simple to
recover the Lindblad rate equation corresponding to the first case, Eq. (\ref%
{primerol}). On the other hand, the second case, Eq. (\ref{segundol}),
follows by defining the contribution $\mathcal{C}_{sa}$ as%
\begin{equation}
\mathcal{C}_{sa}[\rho ]=\frac{1}{2}\sum_{\alpha }\gamma _{\alpha
}([T_{\alpha 0},\rho T_{\alpha 0}^{\dag }]+[T_{\alpha 0}\rho ,T_{\alpha
0}^{\dag }]),  \label{LindbladInteraction2}
\end{equation}%
where the operators $\{T_{\alpha 0}\}$\ are%
\begin{equation}
T_{\alpha 0}=V_{\alpha }\otimes \left\vert a_{\mathrm{0}}\right\rangle
\left\langle a_{\mathrm{0}}\right\vert .  \label{Operators}
\end{equation}%
Notice that in this case [Eq. (\ref{LindbladInteraction2})] an addition over
the ancilla states [Eq. (\ref{LindbladInteraction1})] is not necessary.

\subsection{Non-Markovian system dynamics}

By using projector techniques \cite{breuerbook,haake}, from the bipartite
reformulation [Eq. (\ref{LindbladBipartito})] it is possible to obtain the
system density matrix evolution. Let introduce the projectors $\mathcal{P}$
and $\mathcal{Q}$,%
\begin{equation}
\mathcal{P}\rho _{t}^{sa}=\mathrm{Tr}_{a}\left[ \rho _{t}^{sa}\right]
\otimes \left\vert a_{\mathrm{0}}\right\rangle \left\langle a_{\mathrm{0}%
}\right\vert ,\ \ \ \ \ \ \ \ \mathcal{P}+\mathcal{Q}=\mathrm{I}_{sa},
\label{projectors}
\end{equation}%
where $\mathrm{I}_{sa}$ is the identity matrix in the bipartite
system-ancilla Hilbert space. As usual \cite{breuerbook,haake}, the
bipartite evolution (\ref{LindbladBipartito}) can be projected in relevant
and irrelevant contributions%
\begin{eqnarray}
\frac{d}{dt}\mathcal{P}\rho _{t}^{sa} &=&\mathcal{PL(P}+\mathcal{Q})\rho
_{t}^{sa},  \label{Relevante} \\
\frac{d}{dt}\mathcal{Q}\rho _{t}^{sa} &=&\mathcal{QL(P}+\mathcal{Q})\rho
_{t}^{sa}.  \label{Irrelevante}
\end{eqnarray}%
On the other hand, as an initial condition we consider a separable bipartite
state%
\begin{equation}
\rho _{0}^{sa}=\rho _{0}^{s}\otimes \Pi _{\mathrm{0}}=\rho _{0}^{s}\otimes
\left\vert a_{\mathrm{0}}\right\rangle \left\langle a_{\mathrm{0}%
}\right\vert ,  \label{CISeparable}
\end{equation}%
where $\rho _{0}^{s}$\ is an arbitrary system state. Hence, the ancilla
begins in the pure state $\Pi _{\mathrm{0}}=\left\vert a_{\mathrm{0}%
}\right\rangle \left\langle a_{\mathrm{0}}\right\vert ,$ which in turn
implies that the initial bath-state is $i=0$ $[p_{0}^{i}=\delta _{i0}$ in
Eq. (\ref{classical})].

Given the initial bipartite state (\ref{CISeparable}), it follows that $%
\mathcal{Q}\rho _{0}^{sa}=0.$ Therefore, Eq. (\ref{Irrelevante}) can be
integrated as $\mathcal{Q}\rho _{t}^{sa}=\int_{0}^{t}dt^{\prime }\exp [%
\mathcal{QL}(t-t^{\prime })]\mathcal{QLP}\rho _{t^{\prime }}^{sa},$ which in
turn, after replacing in Eq. (\ref{Relevante}), leads to the convoluted
evolution \cite{haake}%
\begin{equation}
\frac{d}{dt}\mathcal{P}\rho _{t}^{sa}=\mathcal{PLP}\rho _{t}^{sa}+\mathcal{PL%
}\int_{0}^{t}dt^{\prime }\exp [\mathcal{QL}(t-t^{\prime })]\mathcal{QLP}\rho
_{t^{\prime }}^{sa}.  \label{EvolutionProjectores}
\end{equation}%
An explicit system density matrix evolution can be obtained from this
general expression. Its structure depends on each case.

\subsubsection*{First case}

The bipartite superoperator $\mathcal{L}$ is defined by Eq. (\ref%
{LindbladBipartito}). Taking into account Eq. (\ref{LindbladInteraction1})
it can be written as%
\begin{eqnarray}
\mathcal{L}[\bullet ] &=&(\mathcal{L}_{s}+\mathcal{L}_{a})[\bullet ]-\frac{1%
}{2}\sum\nolimits_{\alpha i}^{^{\prime }}\gamma _{\alpha }\{V_{\alpha
}^{\dag }V_{\alpha }\otimes \Pi _{i},\bullet \}_{+}  \notag \\
&&+\sum\nolimits_{\alpha i}^{^{\prime }}\gamma _{\alpha }V_{\alpha
}\left\langle a_{i}\right\vert \bullet \left\vert a_{i}\right\rangle
V_{\alpha }^{\dag }\otimes \Pi _{i},
\end{eqnarray}%
where $\Pi _{i}\equiv \left\vert a_{i}\right\rangle \left\langle
a_{i}\right\vert ,$ and $\{.,.\}_{+}$ denotes an anticommutator operation.
With this expression it is possible to evaluate all contributions in Eq. (%
\ref{EvolutionProjectores}). We get%
\begin{equation}
\mathcal{PL}[\bullet ]=\{\mathcal{L}_{s}(\mathrm{Tr}_{a}\left[ \bullet %
\right] )+\mathcal{C}_{s}[\sum\nolimits_{i}^{^{\prime }}\left\langle
a_{i}\right\vert \bullet \left\vert a_{i}\right\rangle ]\}\otimes \Pi _{%
\mathrm{0}},
\end{equation}%
where $\mathcal{C}_{s}$ is given by Eq. (\ref{LindbladDefinition}). Hence,
it follows the result $\mathcal{PLP}\rho _{t}^{sa}=\mathcal{L}_{s}\rho
_{t}^{s}\otimes \Pi _{\mathrm{0}}.$ Furthermore, $\mathcal{QLP}\rho
_{t^{\prime }}^{sa}=\rho _{t^{\prime }}^{s}\otimes \mathcal{L}_{a}[\Pi _{%
\mathrm{0}}].$ Using the classicality of the ancilla dynamics, Eq. (\ref%
{LindbladClasico}), $\left\langle a_{i}\right\vert \mathcal{L}_{a}[\rho
_{t}]\left\vert a_{i}\right\rangle =\left\langle a_{i}\right\vert \mathcal{L}%
_{a}[\sum_{j}\left\langle a_{j}\right\vert \rho _{t}\left\vert
a_{j}\right\rangle \otimes \Pi _{j}]\left\vert a_{i}\right\rangle $
(populations are only coupled to populations), it is also possible to obtain 
$(\mathcal{QL})^{n}(\rho _{t^{\prime }}^{s}\otimes \mathcal{L}_{a}[\Pi _{%
\mathrm{0}}])=(\mathcal{L}_{s}+\mathcal{C}_{s}+\mathcal{L}_{a})^{n}(\rho
_{t^{\prime }}^{s}\otimes \mathcal{L}_{a}[\Pi _{\mathrm{0}}]),$ with $n=1.$
The validity of this expression for $n=2,3,\cdots $ can be demonstrated by
using the mathematical principle of induction. Hence, we get%
\begin{equation}
\exp [\mathcal{QL}t]\mathcal{QLP}\rho _{t^{\prime }}^{sa}=\exp [(\mathcal{L}%
_{s}+\mathcal{C}_{s}+\mathcal{L}_{a})t](\rho _{t^{\prime }}^{s}\otimes 
\mathcal{L}_{a}[\Pi _{\mathrm{0}}]).
\end{equation}%
After introducing the previous results in Eq. (\ref{EvolutionProjectores})
we obtain the non-Markovian master equation%
\begin{equation}
\frac{d}{dt}\rho _{t}^{s}\!=\!\mathcal{L}_{s}\rho _{t}^{s}+\mathcal{C}%
_{s}\int_{0}^{t}dt^{\prime }k_{\mathrm{I}}(t-t^{\prime })\{\exp
[(t-t^{\prime })(\mathcal{L}_{s}+\mathcal{C}_{s})]\rho _{t^{\prime }}^{s}\},
\label{NoMarkovMasterUno}
\end{equation}%
where the kernel function is defined in terms of the ancilla dynamics 
\begin{subequations}
\label{LindbladKernel1}
\begin{eqnarray}
k_{\mathrm{I}}(t) &=&\sum\nolimits_{i}^{^{\prime }}\left\langle
a_{i}\right\vert \exp (t\mathcal{L}_{a})\mathcal{L}_{a}[\Pi _{\mathrm{0}%
}]\left\vert a_{i}\right\rangle , \\
&=&\frac{d}{dt}\sum\nolimits_{i}^{^{\prime }}\left\langle a_{i}\right\vert
\exp (t\mathcal{L}_{a})[\Pi _{\mathrm{0}}]\left\vert a_{i}\right\rangle .
\end{eqnarray}%
The trace preservation condition $\sum_{i}\left\langle a_{i}\right\vert 
\mathcal{L}_{a}(\bullet )\left\vert a_{i}\right\rangle =0$ is consistent
with $(d/dt)\sum\nolimits_{i}p_{t}^{i}=0.$ Therefore, $\sum_{i}\left\langle
a_{i}\right\vert \mathcal{L}_{a}(\bullet )\left\vert a_{i}\right\rangle
=\left\langle a_{\mathrm{0}}\right\vert \mathcal{L}_{a}(\bullet )\left\vert
a_{\mathrm{0}}\right\rangle +\sum_{i}^{\prime }\left\langle a_{i}\right\vert 
\mathcal{L}_{a}(\bullet )\left\vert a_{i}\right\rangle ,$ which leads to the
equivalent expression 
\end{subequations}
\begin{equation}
k_{\mathrm{I}}(t)=-\frac{d}{dt}\left\langle a_{\mathrm{0}}\right\vert \exp (t%
\mathcal{L}_{a})[\Pi _{\mathrm{0}}]\left\vert a_{\mathrm{0}}\right\rangle =-%
\frac{d}{dt}p_{t}^{0}.  \label{ShortKernel1}
\end{equation}

Eq. (\ref{NoMarkovMasterUno}) is one of the main results of this section. It
provides a natural extension of Shabani-Lidar proposal. In fact, when the
condition $[\mathcal{L}_{s},\mathcal{C}_{s}]=0$ is satisfied, in an
\textquotedblleft interaction representation\textquotedblright\ with respect
to $\mathcal{L}_{s}$ it follows Eq. (\ref{LidarEquation}) with $%
k(t)\rightarrow k_{\mathrm{I}}(t).$ By construction (partial trace over a
Lindblad dynamics) Eq. (\ref{NoMarkovMasterUno}) is a completely positive
evolution, generalizing in this way the results of Ref. \cite{unMedio}.
Notice that after defining the \textquotedblleft
system-reservoir\textquotedblright\ dynamics [Eq. (\ref{primerol})], we have
not introduced any extra approximation in the derivation of this result. On
the other hand, in the present approach the kernel function is completely
determined by the dynamics of the classical environment fluctuations. In
fact, given the initial condition Eq. (\ref{CISeparable}), in Eq. (\ref%
{ShortKernel1}) $p_{t}^{0}$ corresponds to the survival probability of the $%
i=0$ bath-state $[p_{0}^{i}=\delta _{i0}]$ that follows from Eq. (\ref%
{classical}). Notice that any kernel $k_{\mathrm{I}}(t)$ arising from this
classical structure guarantees the completely positive condition of the
solution map, $\rho _{0}^{s}\rightarrow \rho _{t}^{s}.$ If the bath
transitions rates depend explicitly on time, $\{\phi _{ij}\}\rightarrow
\{\phi _{ij}(t)\},$ the kernel becomes non-stationary, $k_{\mathrm{I}%
}(t)\rightarrow k_{\mathrm{I}}(\tau ,t)$ \cite{Lidar}. The corresponding
master equation can be worked out in a similar way.

\subsubsection*{Second case}

In the second case, taking into account the superoperator (\ref%
{LindbladInteraction2}), the bipartite superoperator $\mathcal{L}$ [Eq. (\ref%
{LindbladBipartito})] reads%
\begin{eqnarray}
\mathcal{L}[\bullet ] &=&(\mathcal{L}_{s}+\mathcal{L}_{a})[\bullet ]-\frac{1%
}{2}\sum_{\alpha }\gamma _{\alpha }\{V_{\alpha }^{\dag }V_{\alpha }\otimes
\Pi _{\mathrm{0}},\bullet \}_{+}  \notag \\
&&+\sum_{\alpha }\gamma _{\alpha }V_{\alpha }\left\langle a_{\mathrm{0}%
}\right\vert \bullet \left\vert a_{\mathrm{0}}\right\rangle V_{\alpha
}^{\dag }\otimes \Pi _{\mathrm{0}}.
\end{eqnarray}%
From here, we obtain%
\begin{equation}
\mathcal{PL}[\bullet ]=\{\mathcal{L}_{s}(\mathrm{Tr}_{a}\left[ \bullet %
\right] )+\mathcal{C}_{s}[\left\langle a_{\mathrm{0}}\right\vert \bullet
\left\vert a_{\mathrm{0}}\right\rangle ]\}\otimes \Pi _{\mathrm{0}}.
\end{equation}%
Hence, it follows $\mathcal{PLP}\rho _{t}^{sa}=(\mathcal{L}_{s}+\mathcal{C}%
_{s})\rho _{t}^{s}\otimes \Pi _{\mathrm{0}},$ and $\mathcal{QLP}\rho
_{t^{\prime }}^{sa}=\rho _{t^{\prime }}^{s}\otimes \mathcal{L}_{a}[\Pi _{%
\mathrm{0}}].$ As in the previous case, using the classicality of $\mathcal{L%
}_{a}$ it is also possible to obtain $\mathcal{QL}(\rho _{t^{\prime
}}^{s}\otimes \mathcal{L}_{a}[\Pi _{\mathrm{0}}])=(\mathcal{L}_{s}+\mathcal{L%
}_{a})(\rho _{t^{\prime }}^{s}\otimes \mathcal{L}_{a}[\Pi _{\mathrm{0}}]),$
which by induction leads to%
\begin{equation}
\exp [\mathcal{QL}t]\mathcal{QLP}\rho _{t^{\prime }}^{sa}=\exp [(\mathcal{L}%
_{s}+\mathcal{L}_{a})t](\rho _{t^{\prime }}^{s}\otimes \mathcal{L}_{a}[\Pi _{%
\mathrm{0}}]).
\end{equation}%
By introducing these results in Eq. (\ref{EvolutionProjectores}) we get%
\begin{equation*}
\frac{d}{dt}\rho _{t}^{s}=(\mathcal{L}_{s}+\mathcal{C}_{s})\rho _{t}^{s}-%
\mathcal{C}_{s}\int_{0}^{t}dt^{\prime }k_{\mathrm{I}}(t-t^{\prime })\{\exp
[(t-t^{\prime })\mathcal{L}_{s}]\rho _{t^{\prime }}^{s}\},
\end{equation*}%
where $k_{\mathrm{I}}(t)$ is given by Eq. (\ref{ShortKernel1}). This master
equation can trivially be rewritten as%
\begin{equation}
\frac{d}{dt}\rho _{t}^{s}=\mathcal{L}_{s}\rho _{t}^{s}+\mathcal{C}%
_{s}\int_{0}^{t}dt^{\prime }k_{\mathrm{II}}(t-t^{\prime })\{\exp
[(t-t^{\prime })\mathcal{L}_{s}]\rho _{t^{\prime }}^{s}\}.
\label{MasterSimpler}
\end{equation}%
where the kernel function is $k_{\mathrm{II}}(t)=\delta (t)-k_{\mathrm{I}%
}(t).$ Hence, 
\begin{subequations}
\label{LindbladKernel2}
\begin{eqnarray}
k_{\mathrm{II}}(t) &=&\delta (t)+\frac{d}{dt}\left\langle a_{\mathrm{0}%
}\right\vert \exp (t\mathcal{L}_{a})[\Pi _{\mathrm{0}}]\left\vert a_{\mathrm{%
0}}\right\rangle , \\
&=&\delta (t)+\frac{d}{dt}p_{t}^{0}.
\end{eqnarray}%
If $[\mathcal{L}_{s},\mathcal{C}_{s}]=0,$ in an interaction representation
with respect to $\mathcal{L}_{s},$ from Eq. (\ref{MasterSimpler}) it follows
Eq. (\ref{OneKernel}) with $k(t)\rightarrow k_{\mathrm{II}}(t).$ Therefore,
that equation is obtained, without introducing any approximation, from the
alternative system-environment coupling defined by Eq. (\ref{segundol}).
This is the second main result of this section.

Contrary to the previous case [Eq. (\ref{ShortKernel1})], in this one the
kernel includes a delta term, which in turn leads to a local in time
contribution in the evolution (\ref{MasterSimpler}). The presence of this
term can be explained directly from Eq. (\ref{segundol}). In fact, in the
limit where the bath does not fluctuate, $\phi _{ij}\rightarrow 0,$ taking
into account the initial condition (\ref{CISeparable}), a Markovian Lindblad
dynamics defined by $(\mathcal{L}_{s}+\mathcal{C}_{s})$ is recovered.

When the condition (\ref{NoColision}) is not satisfied, Eq. (\ref%
{MasterSimpler}) can also be derived from a different underlying dynamics
that gives an alternative expression for the memory kernel \cite{Embedding}.
As the present approach relies on condition (\ref{NoColision}), it provides
an alternative basis for the derivation of Eq. (\ref{MasterSimpler}), which
in consequence can be applied in a larger range of dissipative dynamics.

\subsection{Exponential kernels}

Different analysis of Eqs. (\ref{LidarEquation}) and (\ref{OneKernel}) were
performed after assuming an exponential kernel \cite%
{ManisPetru,unMedio,piilo}. In the present approach, this particular case
arises when the environment has only two different configurational states,
that is, the ancilla Hilbert space is defined by only two states, $%
\left\vert a_{\mathrm{0}}\right\rangle $ and $\left\vert a_{1}\right\rangle
. $ The classical master equation (\ref{classical}) becomes 
\end{subequations}
\begin{equation}
\frac{d}{dt}p_{t}^{0}=-\phi p_{t}^{0}+\varphi p_{t}^{1},\ \ \ \ \ \ \ \ \ \
\ \frac{d}{dt}p_{t}^{1}=-\varphi p_{t}^{1}+\phi p_{t}^{0}.
\label{MasterTLSEXPO}
\end{equation}%
Here, the bath transition rates are denoted by $\phi $\ and $\varphi .$ For
normalized initial conditions $p_{0}^{0}+p_{0}^{1}=1,$ with $p_{0}^{0}=1,$
and $p_{0}^{1}=0$ [Eq. (\ref{CISeparable})] the solutions are 
\begin{subequations}
\label{ClassicalTLS}
\begin{eqnarray}
p_{t}^{0} &=&\frac{\varphi }{\phi +\varphi }+\exp [-t(\phi +\varphi )]\frac{%
\phi }{\phi +\varphi }, \\
p_{t}^{1} &=&\frac{\phi }{\phi +\varphi }-\exp [-t(\phi +\varphi )]\frac{%
\phi }{\phi +\varphi }.
\end{eqnarray}%
Thus, the kernel (\ref{ShortKernel1}) reads 
\end{subequations}
\begin{equation}
k_{\mathrm{I}}(t)=\phi \exp [-t(\phi +\varphi )],  \label{kernelExpor}
\end{equation}%
while in the second case, Eq. (\ref{LindbladKernel2}), it follows%
\begin{equation}
k_{\mathrm{II}}(t)=\delta (t)-\phi \exp [-t(\phi +\varphi )].
\label{kernelExporDelta}
\end{equation}%
In this way, our formalism associates a very simple environment structure to
these kinds of kernels. Notice that in the second case, Eq. (\ref%
{kernelExporDelta}), the exponential contribution is accompanied by a local
in time contribution. In fact, as well known \cite{stenholm,classBu}, Eq. (%
\ref{OneKernel}) with a single exponential kernel does not lead in general
to a completely positive dynamics \cite{ManisPetru,unMedio}. Hence, the
delta contribution avoids the lack of the completely positive condition of
the solution map. On the other hand, both kernels, as well as the underlying
physical dynamics, are well defined for any value of the parameters. In
particular when $\varphi =0.$

\subsection{Extension to non-interacting bipartite systems}

In Ref. \cite{grano} the authors analyzed a generalization of Eq. (\ref%
{LidarEquation}) for two systems, where one of them obeys a unitary
evolution. That situation can be easily described in the present frame.
Apart from the system $S,$ we consider another system $S^{\prime }.$ The
single evolution of $S^{\prime }$ is local in time (Markovian) and defined
by a superoperator $\mathcal{L}_{s^{\prime }}.$ Then, we ask about the
evolution of the bipartite density matrix $\rho _{t}^{ss^{\prime }}$
associated to both systems. This question can be straightforwardly answered
from Eq. (\ref{NoMarkovMasterUno}) by taking as system the bipartite
extension $S\otimes S^{\prime }.$ Therefore, under the replacements $\rho
_{t}^{s}\rightarrow \rho _{t}^{ss^{\prime }}$ and $\mathcal{L}%
_{s}\rightarrow (\mathcal{L}_{s}+\mathcal{L}_{s^{\prime }}),$ $\mathcal{C}%
_{s}\rightarrow \mathcal{C}_{s}\otimes \mathrm{I}_{s^{\prime }},$\ where $%
\mathrm{I}_{s^{\prime }}$ is the identity matrix, it follows%
\begin{equation}
\frac{d}{dt}\rho _{t}^{ss^{\prime }}=(\mathcal{L}_{s}+\mathcal{L}_{s^{\prime
}})\rho _{t}^{ss^{\prime }}+\mathcal{C}_{s}\int_{0}^{t}dt^{\prime }k_{%
\mathrm{I}}(t^{\prime })e^{t^{\prime }(\mathcal{L}_{s}+\mathcal{L}%
_{s^{\prime }}+\mathcal{C}_{s})}\rho _{t-t^{\prime }}^{ss^{\prime }}.
\label{bipartirol}
\end{equation}%
Assuming a separable initial bipartite state, $\rho _{0}^{ss^{\prime }}=\rho
_{0}^{s}\otimes $ $\rho _{0}^{s^{\prime }},$ the solution of Eq. (\ref%
{bipartirol}) can be written as $\rho _{t}^{ss^{\prime }}=\rho
_{t}^{s}\otimes \rho _{t}^{s^{\prime }},$ where $\rho _{t}^{s}$ obey Eq. (%
\ref{NoMarkovMasterUno}), while the evolution of $\rho _{t}^{s^{\prime }}$
is local in time and defined by the superoperator $\mathcal{L}_{s^{\prime
}}, $ that is, $(d/dt)\rho _{t}^{s^{\prime }}=\mathcal{L}_{s^{\prime }}\rho
_{t}^{s^{\prime }}.$ The evolution proposed in Ref. \cite{grano} is
recovered by taking $[\mathcal{L}_{s},\mathcal{C}_{s}]=0$ in Eq. (\ref%
{bipartirol}), that is, when $\exp [t(\mathcal{L}_{s}+\mathcal{C}_{s})]=\exp
[t\mathcal{C}_{s}]\exp [t\mathcal{L}_{s}].$ On the other hand, the same kind
of bipartite extension applies to Eq. (\ref{MasterSimpler}).

\section{Measurement trajectories}

In this section we analyze the situation in which the system is subjected to
a continuous-in-time measurement action. Provided that the joint dynamics of
the system and the environment transitions admits a Markovian representation
in a bipartite Hilbert space, a standard (Markovian) quantum jump approach 
\cite{plenio,carmichaelbook}\ can be formulated for the description of this
problem. In fact, it is possible to define a stochastic density matrix $\rho
_{\mathrm{st}}^{sa}(t)$ whose ensemble of (measurement) realizations
recovers the dynamics of the bipartite state $\rho _{t}^{sa},$ Eq. (\ref%
{LindbladBipartito}), that is, $\overline{\rho _{\mathrm{st}}^{sa}}(t)=\rho
_{t}^{sa},$ where the overbar denotes the ensemble average. Nevertheless, in
general it is not possible to write down a \textquotedblleft closed
dynamics\textquotedblright\ for the realizations projected over the system
Hilbert space \cite{OneChannel}, $\rho _{\mathrm{st}}^{s}(t)=\mathrm{Tr}%
_{a}[\rho _{\mathrm{st}}^{sa}(t)].$ In fact, generally the evolution of $%
\rho _{\mathrm{st}}^{s}(t)$\ cannot be written without involving explicitly
the ancilla state. In the Appendix, \textquotedblleft \textit{over the basis}%
\textquotedblright\ of the bipartite representation, we demonstrate that Eq.
(\ref{NoMarkovMasterUno}) can be unravelled in terms of an ensemble of
realizations with a closed evolution only when the measurement process is a
renewal one\ \cite{barchiellibook,plenio,carmichaelbook,carmichael,bartano}
and the ancilla Hilbert space is bidimensional. On the other hand, for Eq. (%
\ref{MasterSimpler}) the renewal property is also necessary. Nevertheless,
the ancilla Hilbert space may be arbitrary.

Note that the previous conditions rely on two central ingredients, that is,
the system dynamics admits a Markovian representation in a higher Hilbert
space and the system realizations have the same structure than in the
Markovian case \cite{OneChannel}. Hence, they do not depend explicitly on
the configurational bath degrees of freedom. It is possible to conjecture
that more (unknown) general formalisms based on different hypothesis could
lead to different unravelling conditions.

After introducing a set of system operators that lead to renewal measurement
processes, over the basis of the non-Markovian master equations (\ref%
{NoMarkovMasterUno}) and (\ref{MasterSimpler}) we derive their corresponding
ensemble of measurement realizations.

\subsection*{Renewal measurements processes}

In renewal measurement processes, the information about the system state is
completely lost after a detection event \cite%
{barchiellibook,plenio,carmichaelbook,carmichael,bartano}. This property is
induced by the operators $\{V_{\alpha }\}$ that define $\mathcal{C}_{s},$
Eq. (\ref{LindbladDefinition}). We assume that%
\begin{equation}
V_{\alpha }=\left\vert r_{\alpha }\right\rangle \left\langle u\right\vert ,
\label{VRenewal}
\end{equation}%
where $\left\vert r_{\alpha }\right\rangle $ and $\left\vert u\right\rangle $
are system states. For establishing the subsequent notation, we rewrite $%
\mathcal{C}_{s}$ as%
\begin{equation}
\mathcal{C}_{s}=\mathcal{D}_{s}+\mathcal{J}_{s}.  \label{spliting}
\end{equation}%
The superoperator $\mathcal{D}_{s}$ reads%
\begin{equation}
\mathcal{D}_{s}[\rho ]=-\frac{1}{2}\sum_{\alpha }\gamma _{\alpha
}\{V_{\alpha }^{\dag }V_{\alpha },\rho \}_{+}=-\frac{1}{2}\gamma
\{\left\vert u\right\rangle \left\langle u\right\vert ,\rho \}_{+},
\end{equation}%
where the rate is $\gamma =\sum_{\alpha }\gamma _{\alpha }.$ The
\textquotedblleft jump\textquotedblright\ superoperator $\mathcal{J}_{s}$\ is%
\begin{equation}
\mathcal{J}_{s}[\rho ]=\sum_{\alpha }\gamma _{\alpha }V_{\alpha }\rho
V_{\alpha }^{\dag }=\gamma \bar{\rho}_{s}\left\langle u\right\vert \rho
\left\vert u\right\rangle ,
\end{equation}%
where the system \textquotedblleft resetting state\textquotedblright\ $\bar{%
\rho}_{s}$ is%
\begin{equation}
\bar{\rho}_{s}=\sum_{\alpha }p_{\alpha }\left\vert r_{\alpha }\right\rangle
\left\langle r_{\alpha }\right\vert ,\ \ \ \ \ \ \ \ \ \ \ p_{\alpha }=\frac{%
\gamma _{\alpha }}{\sum_{\alpha ^{\prime }}\gamma _{\alpha ^{\prime }}}.
\label{reseting}
\end{equation}%
Given the operators (\ref{VRenewal}), it is simple to realize that $\mathcal{%
J}_{s}$ can be rewritten as%
\begin{equation}
\mathcal{J}_{s}[\rho ]=\bar{\rho}_{s}\mathrm{Tr}_{s}[\mathcal{J}_{s}\rho ].
\label{renewalProperty}
\end{equation}

Assuming that the measurement apparatus is sensitive to all system
transitions $(\left\vert u\right\rangle \rightsquigarrow \left\vert
r_{\alpha }\right\rangle )$\ introduced by the operators $V_{\alpha }$ [Eq. (%
\ref{VRenewal})], the measurement transformation associated to each
detection event reads \cite{breuerbook}%
\begin{equation}
\mathcal{M}[\rho ]=\frac{\mathcal{J}_{s}\rho }{\mathrm{Tr}_{s}[\mathcal{J}%
_{s}\rho ]}=\bar{\rho}_{s}.  \label{measurolino}
\end{equation}%
Therefore, after a detection event all information about the pre-measurement
state is lost, and the system collapses to the state $\bar{\rho}_{s}$ \cite%
{plenio,carmichaelbook,carmichael,bartano}. Below, starting from the
corresponding non-Markovian master equations, we formulate a quantum jump
approach for each case.

\subsubsection*{First case}

The solution of the convoluted evolution (\ref{NoMarkovMasterUno}) can be
written in the Laplace domain $[\hat{f}(u)\equiv \int_{0}^{\infty
}dte^{-ut}f(t)]$ as%
\begin{equation}
\hat{\rho}_{u}^{s}=\mathcal{\hat{G}}(u)[\rho _{0}^{s}]=\frac{1}{u-\mathcal{L}%
_{s}-\mathcal{C}_{s}\hat{k}_{\mathrm{I}}(u-\mathcal{L}_{s}-\mathcal{C}_{s})}%
\rho _{0}^{s}.  \label{Rholu}
\end{equation}%
In agreement with the results of the Appendix, we assume an exponential
kernel, Eq. (\ref{kernelExpor}). Hence,%
\begin{equation}
\hat{k}_{\mathrm{I}}(u)=\frac{\phi }{u+\phi +\varphi }.
\end{equation}%
In Eq. (\ref{Rholu}) we used the consistent notation $\hat{f}(u-\mathcal{O}%
)\equiv \int_{0}^{\infty }dte^{-ut}f(t)\exp (t\mathcal{O}),$ where $\mathcal{%
O}$ is an arbitrary superoperator. For simplifying the next calculations
steps we also introduce the superoperators 
\begin{subequations}
\label{Laplor}
\begin{eqnarray}
\mathcal{\hat{D}}_{s}(u) &=&\mathcal{D}_{s}\hat{k}_{\mathrm{I}}(u-\mathcal{L}%
_{s}-\mathcal{D}_{s}), \\
\mathcal{\hat{J}}_{s}(u) &=&\mathcal{J}_{s}\hat{k}_{\mathrm{I}}(u-\mathcal{L}%
_{s}-\mathcal{D}_{s}).
\end{eqnarray}%
Notice that $\mathcal{\hat{D}}_{s}(u)+\mathcal{\hat{J}}_{s}(u)\neq \mathcal{C%
}_{s}\hat{k}_{\mathrm{I}}(u-\mathcal{L}_{s}-\mathcal{C}_{s}).$ Furthermore,
it is introduced the \textquotedblleft unnormalized conditional
propagator\textquotedblright 
\end{subequations}
\begin{equation}
\mathcal{\hat{T}}(u)=\frac{1}{u-\mathcal{L}_{s}-\mathcal{\hat{D}}_{s}(u)}.
\label{TLapla}
\end{equation}

After some tedious but standard calculations steps, Eq. (\ref{Rholu}) can be
rewritten as%
\begin{equation}
\hat{\rho}_{u}^{s}=\mathcal{\hat{T}}(u)\rho _{0}^{s}+\mathcal{\hat{G}}%
^{\prime }(u)\mathcal{\hat{J}}_{s}(u)\mathcal{\hat{T}}(u)\rho _{0}^{s},
\label{RhoFirstStep}
\end{equation}%
where the additional propagator $\mathcal{\hat{G}}^{\prime }(u)$\ reads%
\begin{equation}
\mathcal{\hat{G}}^{\prime }(u)=\frac{1}{u-\mathcal{L}_{s}-\mathcal{C}_{s}%
\hat{k}_{\mathrm{II}}^{\prime }(u-\mathcal{L}_{s})}.  \label{GPrima}
\end{equation}%
Its associated kernel $\hat{k}_{\mathrm{II}}^{\prime }(u)$ is%
\begin{equation}
\hat{k}_{\mathrm{II}}^{\prime }(u)=1-\frac{\varphi }{u+\phi +\varphi }.
\label{kernelprimado}
\end{equation}%
Therefore, in the time domain it reads\ $k_{\mathrm{II}}^{\prime }(t)=\delta
(t)-\varphi \exp [-t(\phi +\varphi )].$ We remark that Eq. (\ref%
{RhoFirstStep}) was obtained after assuming an exponential kernel. On the
other hand, $\mathcal{\hat{G}}^{\prime }(u)$ corresponds to the propagator
of the evolution (\ref{MasterSimpler}) after interchanging the role of the
environment states $\left\vert 0\right\rangle \leftrightarrow \left\vert
1\right\rangle ,$ which implies the interchange of the corresponding
transition bath rates $\phi \leftrightarrow \varphi .$ This property is
evident in the definition of $\hat{k}_{\mathrm{II}}^{\prime }(u)$ [compare
Eq. (\ref{kernelExporDelta}) in the Laplace domain with Eq. (\ref%
{kernelprimado})]. While the derivation of Eq. (\ref{RhoFirstStep}) is only
valid for exponential kernels, the following calculation steps do not rely
on that assumption.

Using the renewal property (\ref{renewalProperty}), from Eq. (\ref{Laplor})
it follows the property $\mathcal{\hat{J}}_{s}(u)[\rho ]=\bar{\rho}_{s}%
\mathrm{Tr}_{s}[\mathcal{\hat{J}}_{s}(u)\rho ].$ Hence, Eq. (\ref%
{RhoFirstStep}) can be rewritten as%
\begin{equation}
\hat{\rho}_{u}^{s}=\mathcal{\hat{T}}(u)\rho _{0}^{s}+\mathcal{\hat{G}}%
^{\prime }(u)\bar{\rho}_{s}\mathrm{Tr}_{s}[\mathcal{\hat{J}}_{s}(u)\mathcal{%
\hat{T}}(u)\rho _{0}^{s}].  \label{Intermesso}
\end{equation}%
On the other hand, the propagator $\mathcal{\hat{G}}^{\prime }(u)$\ [Eq. (%
\ref{GPrima})] can be rewritten in terms of a series expansion as%
\begin{equation}
\mathcal{\hat{G}}^{\prime }(u)=\mathcal{\hat{T}}^{\prime
}(u)\sum_{n=0}^{\infty }[\mathcal{\hat{J}}_{s}^{\prime }(u)\mathcal{\hat{T}}%
^{\prime }(u)]^{n},  \label{SeriePrima}
\end{equation}%
where the propagator $\mathcal{\hat{T}}^{\prime }(u)$\ is%
\begin{equation}
\mathcal{\hat{T}}^{\prime }(u)=\frac{1}{u-\mathcal{L}_{s}-\mathcal{\hat{D}}%
_{s}^{\prime }(u),}.  \label{TprimaLapla}
\end{equation}%
The superoperators $\mathcal{\hat{D}}_{s}^{\prime }(u)$ and $\mathcal{\hat{J}%
}_{s}^{\prime }(u)$ are 
\begin{subequations}
\begin{eqnarray}
\mathcal{\hat{D}}_{s}^{\prime }(u) &=&\mathcal{D}_{s}\hat{k}_{\mathrm{II}%
}^{\prime }(u-\mathcal{L}_{s}), \\
\mathcal{\hat{J}}_{s}^{\prime }(u) &=&\mathcal{J}_{s}\hat{k}_{\mathrm{II}%
}^{\prime }(u-\mathcal{L}_{s}).
\end{eqnarray}%
From the property (\ref{renewalProperty}), the expansion (\ref{SeriePrima})
becomes 
\end{subequations}
\begin{equation}
\mathcal{\hat{G}}^{\prime }(u)[\bar{\rho}_{s}]=\mathcal{\hat{T}}^{\prime }(u)%
\bar{\rho}_{s}\sum_{n=0}^{\infty }\hat{w}^{n}(u),
\end{equation}%
which introduced in Eq. (\ref{Intermesso}) leads to%
\begin{equation}
\hat{\rho}_{u}^{s}=\mathcal{\hat{T}}(u)\rho _{0}^{s}+\mathcal{\hat{T}}%
^{\prime }(u)\bar{\rho}_{s}\sum_{n=0}^{\infty }\hat{w}^{n}(u)\hat{w}_{%
\mathrm{in}}(u).  \label{finalLaplace}
\end{equation}%
Here, the \textquotedblleft waiting time probability
density\textquotedblright\ $\hat{w}(u)$ is%
\begin{equation}
\hat{w}(u)=\mathrm{Tr}_{s}[\mathcal{\hat{J}}_{s}^{\prime }(u)\mathcal{\hat{T}%
}^{\prime }(u)\bar{\rho}_{s}],  \label{waiter}
\end{equation}%
where $\bar{\rho}_{s}$ is the resetting state (\ref{reseting}). The
\textquotedblleft initial waiting time probability
density\textquotedblright\ $\hat{w}_{\mathrm{in}}(u)$ is%
\begin{equation}
\hat{w}_{\mathrm{in}}(u)=\mathrm{Tr}_{s}[\mathcal{\hat{J}}_{s}(u)\mathcal{%
\hat{T}}(u)\rho _{0}^{s}].  \label{waiterIn}
\end{equation}%
Finally, from Eq. (\ref{finalLaplace}), by using the resetting property of
the measurement transformation $\mathcal{M}$, Eq. (\ref{measurolino}), the\
system density matrix $\rho _{t}^{s}$ can be written as%
\begin{equation}
\rho _{t}^{s}=\sum_{n=0}^{\infty }\rho _{t}^{(n)},  \label{Sumatoria}
\end{equation}%
where each contribution $\rho _{t}^{(n)}$ reads%
\begin{eqnarray}
\rho _{t}^{(n)}\! &=&\!\int_{0}^{t}dt_{n}\cdots \!\int_{0}^{t_{2}}dt_{1}\
P_{n}[t,\{t_{i}\}_{1}^{n}]  \label{RhoEneEstadistica} \\
&&\times \mathcal{T}_{c}^{\prime }(t-t_{n})\mathcal{M}\cdots \mathcal{T}%
_{c}^{\prime }(t_{2}-t_{1})\mathcal{MT}_{c}(t_{1})\rho _{0}^{s},\ \ \ \ \ \
\   \notag
\end{eqnarray}%
$(n\geq 1)$ and $\rho _{t}^{(0)}=P_{0}^{\mathrm{in}}(t)\mathcal{T}%
_{c}(t)\rho _{0}^{s}.$ The \textquotedblleft conditional normalized
propagators\textquotedblright\ $\mathcal{T}_{c}(t)$ and $\mathcal{T}%
_{c}^{\prime }(t)$ are%
\begin{equation}
\mathcal{T}_{c}(t)[\rho ]=\frac{\mathcal{T}(t)\rho }{\mathrm{Tr}_{s}[%
\mathcal{T}(t)\rho ]},\ \ \ \ \ \ \ \ \ \mathcal{T}_{c}^{\prime }(t)[\rho ]=%
\frac{\mathcal{T}^{\prime }(t)\rho }{\mathrm{Tr}_{s}[\mathcal{T}^{\prime
}(t)\rho ]},  \label{Normalizados}
\end{equation}%
where $\mathcal{T}(t)$ and $\mathcal{T}^{\prime }(t)$ are defined by their
Laplace transform (\ref{TLapla}) and (\ref{TprimaLapla}) respectively.

As in the Markovian case \cite{plenio,carmichaelbook}, Eqs. (\ref{Sumatoria}%
) and (\ref{RhoEneEstadistica}) allow us to associate to the system density
matrix evolution an ensemble of stochastic realizations that can be put in
one-to-one correspondence with the successive detections of the measurement
apparatus. This is the main result of this section. Each state $\rho
_{t}^{(n)}$ corresponds to the realizations with $n$-measurement events up
to time $t.$ In fact, $\rho _{t}^{(n)}$ consists on successive non-unitary
dynamics interrupted by the collapses introduced by $\mathcal{M}.$ Notice
that until the first event the conditional dynamics is given by $\mathcal{T}%
_{c}(t)$\ while in the posteriori intervals it is given by $\mathcal{T}%
_{c}^{\prime }(t).$

In Eq. (\ref{RhoEneEstadistica}), the probability density of the each
realization, $P_{n}[t,\{t_{i}\}],$ is given by%
\begin{equation}
P_{n}[t,\{t_{i}\}]=P_{0}(t-t_{n})\prod_{j=2}^{n}w(t_{j}-t_{j-1})w_{\mathrm{in%
}}(t_{1}),  \label{ProbabilidadConjunta}
\end{equation}%
where $0<t_{1}<\cdots <t_{n}<t,$ can be read as the detection times. The
\textquotedblleft waiting time densities\textquotedblright\ $w(t)$ and $w_{%
\mathrm{in}}(t)$ are defined by Eqs. (\ref{waiter}) and (\ref{waiterIn})
respectively. Eq. (\ref{ProbabilidadConjunta}) explicitly shows the renewal
property of the measurement\ process. $w(t)$ gives the statistics of the
time interval between successive detection events, while $w_{\mathrm{in}}(t)$
gives the statistics of the first time interval. The \textquotedblleft
survival probabilities\textquotedblright\ $P_{0}(t)$ and $P_{0}^{\mathrm{in}%
}(t)$\ give the probability of not happening any measurement event in a
given interval. $P_{0}^{\mathrm{in}}(t)$ only applies in the first detection
interval, $P_{n}[t,\{t_{i}\}]|_{n=0}=P_{0}^{\mathrm{in}}(t).$ They read%
\begin{equation}
P_{0}(t)=\mathrm{Tr}_{s}[\mathcal{T}^{\prime }(t)\bar{\rho}_{s}],\ \ \ \ \ \
\ \ \ \ P_{0}^{\mathrm{in}}(t)=\mathrm{Tr}_{s}[\mathcal{T}(t)\rho _{0}^{s}].
\label{Survivilinas}
\end{equation}%
Consistently, it is possible to check that the relations $%
(d/dt)P_{0}(t)=-w(t)$ and $(d/dt)P_{0}^{\mathrm{in}}(t)=-w_{\mathrm{in}}(t)$
are fulfilled.

As well known \cite{plenio,OneChannel}, the survival probabilities (\ref%
{Survivilinas}) allow us to formulate an explicit algorithm for generating
the stochastic system state $\rho _{\mathrm{st}}^{s}(t).$ Its average over
realizations recovers the system density matrix,%
\begin{equation}
\rho _{t}^{s}=\overline{\rho _{\mathrm{st}}^{s}}(t).  \label{RhoEstocastica}
\end{equation}%
The first time interval follows from $P_{0}^{\mathrm{in}}(t_{1})=r,$ while
the successive random intervals\ follow by solving $P_{0}(t)=r,$ where $r$\
is a random number in the interval $(0,1).$ The conditional dynamics between
the initial time $(t=0)$ and the first event at time $t_{1}$ is given by $%
\mathcal{T}_{c}(t)$ [Eq. (\ref{Normalizados})]. Notice that the unnormalized
propagator $\mathcal{T}(t)$ also defines the probability density of the
first event, $w_{\mathrm{in}}(t_{1}).$ The posterior conditional dynamics
between successive events is given by $\mathcal{T}_{c}^{\prime }(t)$ [Eq. (%
\ref{Normalizados})].\ Their probability density, $w(t),$ is also defined by
the propagator $\mathcal{T}^{\prime }(t).$ Each time interval ends with the
abrupt collapse defined by $\mathcal{M},$ Eq. (\ref{measurolino}).

The change of propagator $\mathcal{T}(t)\rightarrow \mathcal{T}^{\prime }(t)$
in the previous algorithm can be easily understood from the underlying
Lindblad rate evolution Eq. (\ref{primerol}). In fact, given the initial
condition (\ref{CISeparable}), after the first event the dynamics is
completely equivalent to that defined by Eq. (\ref{segundol}) after
interchanging the role of the environment (ancilla) states $\left\vert
0\right\rangle \leftrightarrow \left\vert 1\right\rangle .$ Consistently,
notice that $\mathcal{T}^{\prime }(t)$ arises from $\mathcal{G}^{\prime
}(t), $ which in fact corresponds to the propagator of the evolution (\ref%
{MasterSimpler}) under the previous interchange. We remark that these
dynamical features are not covered by the formalism developed in Ref. \cite%
{OneChannel}.

\subsubsection*{Second case}

The solution of the second non-Markovian master evolution (\ref%
{MasterSimpler}) reads%
\begin{equation}
\hat{\rho}_{u}^{s}=\mathcal{\hat{G}}(u)[\rho _{0}^{s}]=\frac{1}{u-\mathcal{L}%
_{s}-\mathcal{C}_{s}\hat{k}_{\mathrm{II}}(u-\mathcal{L}_{s})}\rho _{0}^{s}.
\label{solution2}
\end{equation}%
As demonstrated in the Appendix, in this case the formulation of a
\textquotedblleft closed\textquotedblright\ quantum jump approach is valid
for arbitrary kernels $\hat{k}_{\mathrm{II}}(u),$ or equivalently, for an
arbitrary number of environment states.

Similarly to the previous case, we introduce the superoperators%
\begin{equation}
\mathcal{\hat{D}}_{s}(u)=\mathcal{D}_{s}\hat{k}_{\mathrm{II}}(u-\mathcal{L}%
_{s}),\ \ \ \ \ \ \ \ \mathcal{\hat{J}}_{s}(u)=\mathcal{J}_{s}\hat{k}_{%
\mathrm{II}}(u-\mathcal{L}_{s}),  \label{D_JSecond}
\end{equation}%
and the unnormalized conditional propagator%
\begin{equation}
\mathcal{\hat{T}}(u)=\frac{1}{u-\mathcal{L}_{s}-\mathcal{\hat{D}}_{s}(u)}.
\label{TSecond}
\end{equation}%
Notice that here the relation $\mathcal{\hat{D}}_{s}(u)+\mathcal{\hat{J}}%
_{s}(u)=\mathcal{C}_{s}\hat{k}_{\mathrm{II}}(u-\mathcal{L}_{s})$ is
fulfilled. By writing the solution (\ref{solution2}) as $\hat{\rho}_{u}^{s}=[%
\mathcal{\hat{G}}(u)\mathcal{\hat{T}}^{-1}(u)]\mathcal{\hat{T}}(u)\rho
_{0}^{s},$ and using that $\mathcal{\hat{G}}(u)\mathcal{\hat{T}}^{-1}(u)=1+%
\mathcal{\hat{G}}(u)\mathcal{\hat{J}}_{s}(u),$ we get%
\begin{equation}
\hat{\rho}_{u}^{s}=\mathcal{\hat{T}}(u)\rho _{0}^{s}+\mathcal{\hat{G}}(u)%
\mathcal{\hat{J}}_{s}(u)\mathcal{\hat{T}}(u)\rho _{0}^{s}.
\end{equation}%
Under the replacement $\mathcal{\hat{G}}^{\prime }(u)\rightarrow \mathcal{%
\hat{G}}(u)$ Eq. (\ref{RhoFirstStep}) is equivalent to this expression. It
is not difficult to check that all posterior calculations to that equation
are valid in the present case under the replacements $\mathcal{\hat{T}}%
^{\prime }(u)\rightarrow \mathcal{\hat{T}}(u),$ Eq. (\ref{TSecond}), $%
\mathcal{\hat{D}}_{s}^{\prime }(u)\rightarrow \mathcal{\hat{D}}_{s}(u),$ $%
\mathcal{\hat{J}}_{s}^{\prime }(u)\rightarrow \mathcal{\hat{J}}_{s}(u),$ Eq.
(\ref{D_JSecond}), and $\hat{k}_{\mathrm{II}}^{\prime }(u)\rightarrow \hat{k}%
_{\mathrm{II}}(u),$ where $\hat{k}_{\mathrm{II}}(u)$ is an arbitrary kernel
that has the structure (\ref{LindbladKernel2}). Therefore, under the
previous replacements, the ensemble of realizations characterized by Eqs. (%
\ref{Sumatoria}) to (\ref{RhoEstocastica}) unravels the non-Markovian
dynamics defined by Eq. (\ref{MasterSimpler}) (renewal measurement
processes).

The solution (\ref{solution2}) jointly with the definitions (\ref{D_JSecond}%
) allows us to write $u\hat{\rho}_{u}^{s}-\rho _{0}^{s}=[\mathcal{L}_{s}+%
\mathcal{\hat{D}}_{s}(u)+\mathcal{\hat{J}}_{s}(u)]\hat{\rho}_{u}^{s}.$ By
using the property (\ref{renewalProperty}), which leads to $\mathcal{\hat{J}}%
_{s}(u)[\rho ]=\bar{\rho}_{s}\mathrm{Tr}_{s}[\mathcal{\hat{J}}_{s}(u)\rho ],$
the density matrix evolution, in the time domain, can be rewritten as%
\begin{equation*}
\frac{d\rho _{t}^{s}}{dt}=\mathcal{L}_{s}\rho
_{t}^{s}+\int_{0}^{t}\!\!dt^{\prime }\mathcal{D}_{s}(t-t^{\prime })\rho
_{t^{\prime }}^{s}-\bar{\rho}_{s}\int_{0}^{t}\!\!dt^{\prime }\mathrm{Tr}_{s}[%
\mathcal{D}_{s}(t-t^{\prime })\rho _{t^{\prime }}^{s}].
\end{equation*}%
This evolution has the structure predicted in Ref. \cite{OneChannel} for
non-Markovian renewal measurement processes. Notice that Eq. (\ref%
{NoMarkovMasterUno}), even with an exponential kernel cannot be rewritten
with this structure. This feature follows from the change of conditional
evolution described previously.

\section{Example}

Here, we study a dynamics that shows the main features of the developed
approach. As a system we consider a two-level optical transition whose
Hamiltonian reads $H_{s}=\hbar \omega _{s}\sigma _{z}/2,$ where $\omega _{s}$
is the transition frequency between its eigenstates, denoted as $\left\vert
\pm \right\rangle ,$ while $\sigma _{z}$\ is the $z$-Pauli matrix. The
system is coupled to an external resonant laser field \cite{breuerbook}. On
the other hand, the environment fluctuates between two configurational
states. Hence, the ancilla also is a two-level system, whose states are
denoted as $\{\left\vert a_{\mathrm{0}}\right\rangle ,\left\vert
a_{1}\right\rangle \}.$

The system decay is conditioned to the bath-state, Fig.~1. Both dynamics
studied in the previous sections are considered. In the first case, Fig.
1(a), the system decay is inhibited when the bath is in the state $%
\left\vert a_{\mathrm{0}}\right\rangle ,$ while in the second case, Fig.
1(b), it is inhibited in the state $\left\vert a_{1}\right\rangle .$

In an interaction representation with respect to $H_{s}$ the evolution of
the bipartite state $\rho _{t}^{sa}$ [Eq. (\ref{LindbladBipartito})] reads%
\begin{eqnarray}
\frac{d\rho _{t}^{sa}}{dt} &=&\frac{-i\Omega }{2}[\sigma _{x}\otimes \mathrm{%
I}_{a},\rho _{t}^{sa}]+\frac{\gamma }{2}([T,\rho _{t}^{sa}T^{\dagger
}]+[T\rho _{t}^{sa},T^{\dagger }])  \notag \\
&&+\frac{\phi }{2}([A,\rho _{t}^{sa}A^{\dagger }]+[A\rho
_{t}^{sa},A^{\dagger }])  \label{fluor} \\
&&+\frac{\varphi }{2}([A^{\dagger },\rho _{t}^{sa}A]+[A^{\dagger }\rho
_{t}^{sa},A]).  \notag
\end{eqnarray}%
The first unitary contribution introduces the system-laser coherent
coupling. It is written in terms of the $x$-Pauli matrix $\sigma _{x}.$ $%
\mathrm{I}_{a}$ is the identity matrix in the ancilla Hilbert space. The
(ancilla) operator $A$ reads%
\begin{equation}
A=\mathrm{I}_{s}\otimes \left\vert a_{1}\right\rangle \left\langle a_{%
\mathrm{0}}\right\vert ,
\end{equation}%
where $\mathrm{I}_{s}$ is the system identity matrix. With this definition,
from Eqs. (\ref{auxiliarRho}) and (\ref{populor}) it is simple to check that
in Eq. (\ref{fluor}) the Lindblad contributions proportional to the rates $%
\phi $\ and $\varphi $\ lead to the classical master equation (\ref%
{MasterTLSEXPO}). Hence, they take into account the environment
fluctuations. 
\begin{figure}[tbp]
\includegraphics[bb=51 418 745 640,width=8.75cm]{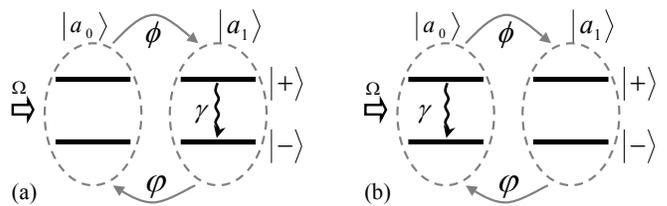}
\caption{Schemes corresponding to the system energy levels and environment
states. (a) First case, Eq. (\protect\ref{primerol}). (b) Second case, Eq. (%
\protect\ref{segundol}). In both schemes, $\Omega $ measures the
system-laser coupling, $\protect\gamma $ is the system decay rate, while $%
\protect\phi $ and $\protect\varphi $\ are the transitions rates between the
two bath-states $\left\vert a_{\mathrm{0}}\right\rangle $ and $\left\vert
a_{1}\right\rangle .$}
\end{figure}

The operator $T$ introduces the natural decay of the system $(\left\vert
+\right\rangle \rightsquigarrow \left\vert -\right\rangle )$ taking into
account its dependence on the bath-states. In the first case [Eq. (\ref%
{primerol})]\ it reads%
\begin{equation}
T=\sigma \otimes \left\vert a_{1}\right\rangle \left\langle a_{1}\right\vert
,
\end{equation}%
while in the second case [Eq. (\ref{segundol})] becomes%
\begin{equation}
T=\sigma \otimes \left\vert a_{\mathrm{0}}\right\rangle \left\langle a_{%
\mathrm{0}}\right\vert .  \label{V_Bipartito}
\end{equation}%
Consistently, the lowering system operator reads $\sigma =\left\vert
-\right\rangle \left\langle +\right\vert .$ The decay rate is $\gamma .$ 
\begin{figure}[tbp]
\includegraphics[bb=20 180 425 550,width=7.5cm]{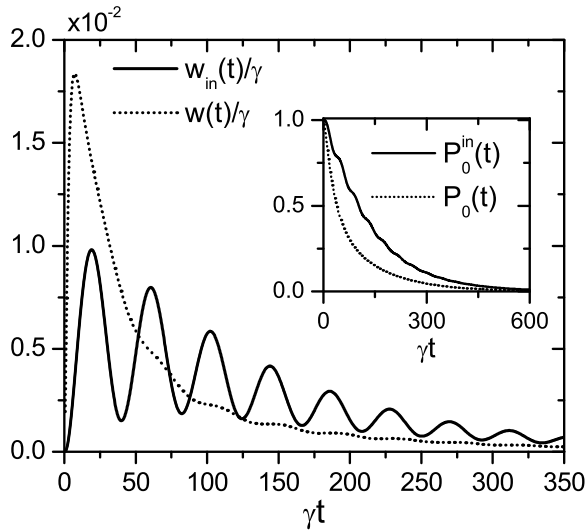}
\caption{Waiting time distributions $w_{\mathrm{in}}(t)$\ [Eq. (\protect\ref%
{waiterIn})] (solid black line) and $w(t)$ [Eq. (\protect\ref{waiter})]
(dotted black line) corresponding to Fig. 1(a) (first case). The inset shows
the associated survival probabilities $P_{0}^{\mathrm{in}}(t)$ and $P_{0}(t)$
respectively, Eq. (\protect\ref{Survivilinas}). The parameters are $\Omega /%
\protect\gamma =0.15,$ and $\protect\phi /\protect\gamma =\protect\varphi /%
\protect\gamma =0.01.$ The initial system state is $\protect\rho %
_{0}^{s}=\left\vert -\right\rangle \left\langle -\right\vert .$ For the same
parameter values, the statistical objects of the second case, Fig. 1(b), are
given by the plots of $w(t)$ and $P_{0}(t).$}
\end{figure}

We remark that in both cases, the evolution defined by Eq. (\ref{fluor})
cannot be mapped with the example worked out in Ref. \cite{OneChannel}. In
fact, although the Lindblad contributions are similar, in that case the
system-ancilla coupling is Hamiltonian. Hence, the ancilla dynamics is
quantum while here it is classical (ancilla populations and coherences are
not coupled).

\subsection{Non-Markovian density matrix evolution}

In agreement with the previous analysis, the initial condition is taken as
[Eq. (\ref{CISeparable})]%
\begin{equation}
\rho _{0}^{sa}=\rho _{0}^{s}\otimes \left\vert a_{\mathrm{0}}\right\rangle
\left\langle a_{\mathrm{0}}\right\vert ,
\end{equation}%
where $\rho _{0}^{s}$\ is an arbitrary system state. Hence, the ancilla
begins in its lower state. Taking into account the previous definitions and
the results of Sec. II it is straightforward to write down the non-Markovian
system density matrix evolution. In the first case, it is given by the
non-Markovian master equation (\ref{NoMarkovMasterUno}) defined with the
exponential kernel (\ref{kernelExpor}), while in the second case, Eq. (\ref%
{MasterSimpler}) with the kernel (\ref{kernelExporDelta}). In both cases the
superoperator $\mathcal{L}_{s}$ and $\mathcal{C}_{s}$ follows from Eq. (\ref%
{fluor}). $\mathcal{L}_{s}$ reads%
\begin{equation}
\mathcal{L}_{s}[\rho ]=-i\frac{\Omega }{2}[\sigma _{x},\rho ],
\end{equation}%
while dissipative effects are introduced by%
\begin{equation}
\mathcal{C}_{s}[\rho ]=\frac{\gamma }{2}([\sigma ,\rho \sigma ^{\dagger
}]+[\sigma \rho ,\sigma ^{\dagger }]).  \label{Cero}
\end{equation}%
By writing $(d/dt)\rho _{t}^{s}=(\mathcal{L}_{s}+\mathcal{C}_{s})\rho
_{t}^{s}$ these superoperators recover the dynamics of a Markovian
fluorescent two-level system \cite{carmichaelbook,breuerbook}. The
dependence of the decay rate $\gamma $ on the bath-states (Fig. 1) introduce
the non-Markovian effects that lead to the evolutions (\ref%
{NoMarkovMasterUno}) and (\ref{MasterSimpler}). Notice that in the first
case the Markovian dynamics is recovered for $\phi \rightarrow \infty ,$ $%
\varphi =0,$ while in the second case for $\phi \rightarrow 0,$ and any $%
\varphi $ (see Fig.~1). It is simple to cheek that these Markovian limits
are achieved by the non-Markovian evolutions (\ref{NoMarkovMasterUno}) and (%
\ref{MasterSimpler}) after taking into account the kernel definitions Eqs. (%
\ref{kernelExpor}) and (\ref{kernelExporDelta}) respectively.

\subsection{Measurement realizations}

In concordance with the dissipative structure (\ref{Cero}), we assume that
the measurement apparatus detects the optical transitions of the system. It
is simple to check that $\mathcal{C}_{s}$ has the renewal structure
corresponding to Eqs. (\ref{VRenewal}) and (\ref{spliting}). The post
measurement state [Eq. (\ref{measurolino})] is $\mathcal{M}[\rho
]=\left\vert -\right\rangle \left\langle -\right\vert .$ Hence, after each
detection event the system collapses to its ground state.

In the first case, the statistics of the time interval between detection
events is defined by the waiting time densities $w_{\mathrm{in}}(t)$\ [Eq. (%
\ref{waiterIn})] and $w(t)$ [Eq. (\ref{waiter})]. In Fig.~2 we plotted these
objects and their associated survival probabilities $P_{0}^{\mathrm{in}}(t)$
and $P_{0}(t)$\ respectively, Eq. (\ref{Survivilinas}). All these
statistical objects can be obtained in an exact analytical way in the
Laplace domain. Nevertheless, contrary to the Markovian case \cite%
{carmichael,bartano}, here the time dependence can only be obtained with
numerical methods. In fact, Laplace inversion via Cauchy's residue theorem,
for arbitrary parameter values, involves roots of a sextic polynomial
(degree 6) in the Laplace variable $u.$ This feature is induced by the
underlying classical transitions of the bath-states, which in turn lead to
dynamical behaviors that depart from the Markovian case.

In the Markovian limit described previously, for $\rho _{0}^{s}=\left\vert
-\right\rangle \left\langle -\right\vert ,$ follows the analytical results $%
w_{\mathrm{in}}(t)=w(t)=4\gamma \Omega ^{2}\exp (-\gamma t/2)[\sinh (\Phi
t/4)/\Phi ]^{2},$ with the \textquotedblleft frequency\textquotedblright\ $%
\Phi \equiv \sqrt{\gamma ^{2}-4\Omega ^{2}}.$ This analytical expression was
obtained previously in Refs. \cite{carmichael,bartano}. For $\Omega
^{2}>\gamma ^{2}/4,$ $w_{\mathrm{in}}(t)$ develops an oscillatory behavior.
Nevertheless, in the non-Markovian case corresponding to Fig.~2, $w_{\mathrm{%
in}}(t)$ develops oscillations even when $\Omega ^{2}<\gamma ^{2}/4.$ This
feature occurs because here the system is able to perform Rabi oscillations
before the first bath transition happens $\left\vert a_{\mathrm{0}%
}\right\rangle \overset{\phi }{\rightarrow }\left\vert a_{1}\right\rangle $
(see Fig. 1), that is, oscillations in $w_{\mathrm{in}}(t)$ appear under the
condition $\Omega >\phi .$ On the other hand, in Fig.~2 $w(t)$ does not
oscillate and approach the Markovian solution \cite{carmichael,bartano}.
This last feature occurs whenever the system is able to perform many optical
transitions before the configurational bath change $\left\vert
a_{1}\right\rangle \overset{\varphi }{\rightarrow }\left\vert a_{\mathrm{0}%
}\right\rangle $ happens. Hence, $w(t)$ approaches the Markovian solution
when $\bar{\tau}^{-1}>\varphi ,$ where $\bar{\tau}$ is the average time
between consecutive emissions in the Markovian case, $\bar{\tau}^{-1}=\gamma
\Omega ^{2}/(\gamma ^{2}+2\Omega ^{2})$ \cite{plenio,carmichaelbook}. For
the parameters of Fig. 2 it follows $\bar{\tau}^{-1}/\varphi \simeq 2.15>1.$
In general, for arbitrary parameters values, $w(t)$ cannot be related to the
waiting time density of the Markovian case.

For the chosen parameter values, $\phi =\varphi ,$ and initial condition, $%
\rho _{0}^{s}=\left\vert -\right\rangle \left\langle -\right\vert ,$ it is
simple to realize that the statistics corresponding to the second case is
completely determined by the waiting time density $w(t)$ and survival
probability $P_{0}(t)$ corresponding to the first case (Fig. 2).%
\begin{figure}[tbp]
\includegraphics[bb=30 260 380 530,width=7.9cm]{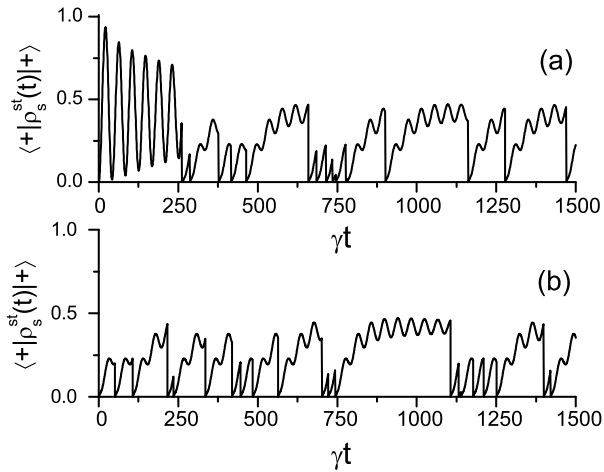}
\caption{Stochastic realizations corresponding to the first (a) and second
case (b). The parameters are the same than in Fig. 2.}
\end{figure}

On the basis of the survival probabilities it is possible to generate the
measurements realizations corresponding to $\rho _{\mathrm{st}}^{s}(t),$ Eq.
(\ref{RhoEstocastica}). In Fig. 3, we show a realization of $\left\langle
+\right\vert \rho _{\mathrm{st}}^{s}(t)\left\vert +\right\rangle .$ Each
detection event corresponds to the collapse to zero of this population. In
agreement with previous analysis, in the first case [Fig. 3(a)] the
conditional dynamics of the first event [Eqs. (\ref{Normalizados}) and (\ref%
{TLapla})] is different of the subsequent ones [Eqs. (\ref{Normalizados})
and (\ref{TprimaLapla})]. In contrast, in the realizations of the second
case [Fig. 3(b)] the conditional dynamics is always the same. Due to the
chosen parameter values $(\phi =\varphi )$ it also corresponds to the
conditional dynamics of the first case after the first event.

In Fig. 4 we plot the upper population $\left\langle +\right\vert \rho
_{t}^{s}\left\vert +\right\rangle $ obtained from the non-Markovian
evolutions Eqs. (\ref{NoMarkovMasterUno}) and (\ref{MasterSimpler}) with the
kernels (\ref{kernelExpor}) and (\ref{kernelExporDelta}) respectively.
Furthermore, we plotted the behavior of $\left\langle +\right\vert \bar{\rho}%
_{\mathrm{st}}^{s}(t)\left\vert +\right\rangle $ obtained by averaging $%
3\times 10^{2}$ realizations shown in Fig. 3. Consistently, in both cases
the master equations fit the average ensemble behavior. This fact shows the
consistence of the non-Markovian quantum jump approach developed in Sec.
III. Furthermore, we checked that the same analytical (Laplace domain) and
numerical results (Fig. 2 to Fig. 4) are obtained by tracing out the
bipartite Markovian representation (see Appendix). 
\begin{figure}[tbp]
\includegraphics[bb=15 0 355 545,width=7.6cm]{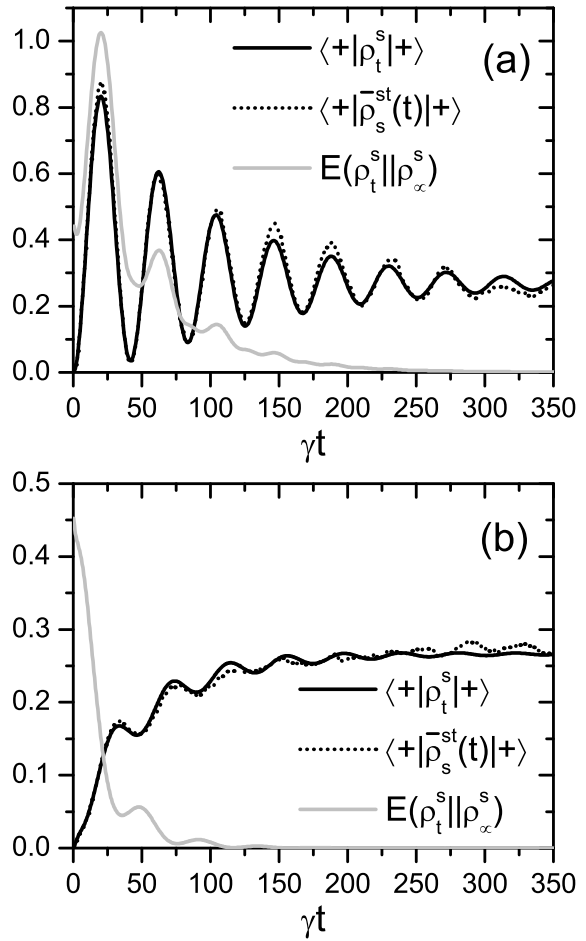}
\caption{Exact solution (black full line) for the upper population $%
\left\langle +\right\vert \protect\rho _{t}^{s}\left\vert +\right\rangle .$
The dotted (noisy) line correspond to an average over $3\times 10^{2}$\
realizations of $\left\langle +\right\vert \protect\rho _{\mathrm{st}%
}^{s}(t)\left\vert +\right\rangle $ shown in Fig. 3. The solid gray line
corresponds to the relative entropy $E(\protect\rho _{t}^{s}||\protect\rho %
_{\infty }^{s}),$ Eq. (\protect\ref{RelativaEstacion}). (a) First case,
defined by the master Eq. (\protect\ref{NoMarkovMasterUno}) with the
exponential kernel (\protect\ref{kernelExpor}). (b) Second case, master Eq. (%
\protect\ref{MasterSimpler}) with the kernel (\protect\ref{kernelExporDelta}%
).}
\end{figure}

\subsection{Environment-to-system backflow of information}

In Ref. \cite{piilo} it was found that both dynamics (\ref{LidarEquation})
and (\ref{OneKernel}) do not lead to \textquotedblleft
genuine\textquotedblright\ non-Markovian effects such as an
environment-to-system backflow of information \cite{NoMeasure}. In the
present approach, that result is completely expectable. In fact, when $[%
\mathcal{L}_{s},\mathcal{C}_{s}]=0,$ from Eqs. (\ref{primerol}) and (\ref%
{segundol}) (take $\mathcal{L}_{s}\rightarrow 0)$ it is simple to realize
that the different environment states only turn-on and turn-off the
Markovian dynamics defined by $\mathcal{C}_{s}.$ Thus, it is impossible to
get an information backflow. In the example of this section that situation
corresponds to take $\Omega =0$ in Fig. 1. Nevertheless, when $[\mathcal{L}%
_{s},\mathcal{C}_{s}]\neq 0,$ the bath fluctuations lead to a switching
between two \textquotedblleft different\textquotedblright\ Markovian
dynamics. Below we show that this underlying effect may lead to a backflow
of information in the generalized dynamics defined by Eqs. (\ref{primerol})
and (\ref{segundol}).

For simplicity, as a witness or detector of the backflow of information we
consider the relative entropy with respect to the stationary state \cite%
{NoMeasure}%
\begin{equation}
E(\rho _{t}^{s}||\rho _{\infty }^{s})=\mathrm{Tr}_{s}[\rho _{t}^{s}(\ln
_{2}\rho _{t}^{s}-\ln _{2}\rho _{\infty }^{s})],  \label{RelativaEstacion}
\end{equation}%
where $\rho _{\infty }^{s}=\lim_{t\rightarrow \infty }\rho _{t}^{s}.$ Hence,
the backflow of information occurs if there exist times $t_{2}>t_{1}$ such
that $E(\rho _{t_{2}}^{s}||\rho _{\infty }^{s})>E(\rho _{t_{1}}^{s}||\rho
_{\infty }^{s})$ \cite{OneChannel,Embedding}.

By working in a Laplace domain, the stationary states corresponding to the
schemes of Fig. 1 can be obtained in an exact way. As $\rho _{\infty }^{s}$
does not depend on the (system-bath) initial condition, it follows the
property $\rho _{\infty }^{s}(\phi ,\varphi )|_{\mathrm{I}}=\rho _{\infty
}^{s}(\varphi ,\phi )|_{\mathrm{II}},$ where $\rho _{\infty }^{s}(\phi
,\varphi )|_{\mathrm{I}}$ is the stationary state of the first case with
parameters $(\phi ,\varphi ),$ while $\rho _{\infty }^{s}(\varphi ,\phi )|_{%
\mathrm{II}}$ is the stationary of the second case where the role of the
parameters is interchanged, that is, $\varphi \leftrightarrow \phi .$

For each case, in Fig. 4 we also plotted the relative entropy (\ref%
{RelativaEstacion}) (solid gray lines). $E(\rho _{t}^{s}||\rho _{\infty
}^{s})$ develops \textquotedblleft revivals\textquotedblright\ showing that
in both cases the dynamics lead to a backflow of information. Consistently
with the analysis of Ref. \cite{piilo} this phenomenon only appears if $%
\Omega \neq 0,$ that is, when the dissipative and unitary contributions do
not commutate.

\section{Summary and conclusions}

In this paper we established a solid physical basis for an alternative
derivation and understanding of two extensively studied non-Markovian master
equations. As underlying \textquotedblleft microscopic
dynamics\textquotedblright\ we utilized a system coupled to an environment
that is able to develop classical self-fluctuations which in turn modify the
system dissipative dynamics, Eqs. (\ref{primerol}) and (\ref{segundol}).
From a bipartite system-ancilla representation, Eq. (\ref{LindbladBipartito}%
), and by means of a projector technique, we obtained the non-Markovian
master equations (\ref{NoMarkovMasterUno}) and (\ref{MasterSimpler}), which
represent one of the main results of this work. If the unitary and
dissipative superoperators commutate, Shabani-Lidar equation, Eq. (\ref%
{LidarEquation}), and its associated evolution, Eq. (\ref{OneKernel}), are
recovered respectively.

In contrast with phenomenological approaches, here the statistical behavior
of the environment fluctuations completely determine the memory functions,
Eqs. (\ref{ShortKernel1}) and (\ref{LindbladKernel2}). The paradigmatic case
of exponential kernels arises when the environment has only two
configurational states, Eqs. (\ref{kernelExpor}) and (\ref{kernelExporDelta}%
). By construction, kernels associated to an arbitrary number of
bath-states, Eq. (\ref{classical}), also guaranty the completely positive
condition of the solution map for any value of the underlying characteristic
parameters.

On the basis of the bipartite representation, we found the conditions under
which the system dynamics can be unravelled in terms of an ensemble of
measurement realizations whose dynamics can be written in a closed way, that
is, without involving explicit information about the configurational
bath-states. Eq. (\ref{NoMarkovMasterUno}) can be unravelled when the bath
has two configurational states, while for Eq. (\ref{MasterSimpler}) this
condition is not necessary. Nevertheless, in both cases a renewal condition
is required. As in the standard Markovian case, the realizations consist of
periods where the evolution is smooth and non-unitary, while at the
detection times the system suddenly collapses to the same resetting state.
The non-Markovian features of the dynamics are present in the conditional
dynamics between jumps, which in turn may be different from the first
interval. The unravelling of the system dynamics in terms of (closed)
measurement trajectories remains as an open problem when the previous
conditions are not satisfied.

The consistence of the previous findings has been explicitly demonstrated by
studying the dynamics of a two-level system whose decay is modulated by the
environment states. This example allowed us to show that a backflow of
information from the environment to the system appears in both master
equations. Therefore, the absence of this phenomenon for the particular
situation analyzed in Ref. \cite{piilo} is not a general property of the
dynamics. In fact, we demonstrated that the backflow of information may
occurs only when the system unitary dynamics does not commutate with the
dissipative one.

In summary, the formalism presented here gives a clear physical
interpretation of some phenomenological aspects that appears in the
formulation of the studied non-Markovian quantum master equations. These
results are of help for understanding and modeling the great variety of
phenomena emerging in presence of memory effects.

\section*{Acknowledgments}

The author thanks to M. Guraya for a critical reading of the manuscript.
This work was supported by CONICET, Argentina, under Grant No. PIP
11420090100211.

\appendix*

\section{Quantum jumps from the bipartite representation}

The properties of the non-Markovian quantum jump approach developed in Sec.\
III can be derived from a standard Markovian quantum jump approach
formulated on the basis of the bipartite dynamics (\ref{LindbladBipartito}).

\subsection{Conditions for getting a closed measurement ensemble}

Firstly, we derive the conditions under which the non-Markovian dynamics (%
\ref{NoMarkovMasterUno}) and (\ref{MasterSimpler}) can be unravelled in
terms of an ensemble of trajectories whose dynamics can be written in a
closed form, that is, without involving in an explicit way the ancilla
(bath) states. As mentioned in Section III, these results rely on the
bipartite representation of the non-Markovian system dynamics.

As demonstrated in Ref. \cite{OneChannel}, a \textquotedblleft closed
quantum jump approach\textquotedblright\ can be formulated for the system of
interest if the bipartite dynamics Eq. (\ref{LindbladBipartito}) fulfill the
conditions%
\begin{equation}
\mathrm{Tr}_{a}[\mathbb{M}\rho _{sa}]=\mathcal{M}[\mathrm{Tr}_{a}(\rho
_{sa})],\ \ \ \ \ \ \ \ \ \ \ \ \mathrm{Tr}_{s}[\mathbb{M}\rho _{sa}]=\bar{%
\rho}_{a}.  \label{ConditionsClosed}
\end{equation}%
Here, $\mathbb{M}$ is the bipartite transformation associated to each
detection event and $\bar{\rho}_{a}$ is an arbitrary ancilla state. The
first condition guarantees that each measurement event in the bipartite
space can also be read as a measurement transformation $\mathcal{M}$ in the
system Hilbert space. The second condition guarantees that the conditional
system dynamics between detection events does not depend explicitly on the
ancilla state \cite{OneChannel}. Therefore, under the previous conditions,
the system measurement dynamics has the same structure (non-unitary
conditional dynamics\ followed by state collapses) than in the Markovian
case.

Assuming that the measurement apparatus is sensitive to all system
transitions defined by the operators $V_{\alpha }$ [Eq. (\ref%
{LindbladDefinition})], it follows \cite{breuerbook}%
\begin{equation}
\mathcal{M}[\rho ]=\frac{\sum\nolimits_{\alpha }\gamma _{\alpha }V_{\alpha
}\rho V_{\alpha }^{\dag }}{\mathrm{Tr}_{s}[\sum\nolimits_{\alpha }\gamma
_{\alpha }V_{\alpha }^{\dag }V_{\alpha }\rho ]}.
\end{equation}%
In the first case, from Eqs. (\ref{LindbladBipartito}) and (\ref%
{LindbladInteraction1}), the bipartite transformation $\mathbb{M}$ reads%
\begin{equation}
\mathbb{M}[\rho ]=\frac{\sum\nolimits_{\alpha ,i}^{^{\prime }}\gamma
_{\alpha }T_{\alpha i}\rho T_{\alpha i}^{\dag }}{\mathrm{Tr}%
_{sa}[\sum\nolimits_{\alpha ,i}^{^{\prime }}\gamma _{\alpha }T_{\alpha
i}^{\dag }T_{\alpha i}\rho ]},  \label{M1}
\end{equation}%
where $T_{\alpha i}=V_{\alpha }\otimes \left\vert a_{i}\right\rangle
\left\langle a_{i}\right\vert ,$ Eq. (\ref{TAlfa_i}). On the other hand, in
the second case, from Eqs. (\ref{LindbladBipartito}) and (\ref%
{LindbladInteraction2}), it becomes%
\begin{equation}
\mathbb{M}[\rho ]=\frac{\sum\nolimits_{\alpha }\gamma _{\alpha }T_{\alpha
0}\rho T_{\alpha 0}^{\dag }}{\mathrm{Tr}_{sa}[\sum\nolimits_{\alpha }\gamma
_{\alpha }T_{\alpha 0}^{\dag }T_{\alpha 0}\rho ]},  \label{M2}
\end{equation}%
where $T_{\alpha 0}=V_{\alpha }\otimes \left\vert a_{\mathrm{0}%
}\right\rangle \left\langle a_{\mathrm{0}}\right\vert ,$ Eq. (\ref{Operators}%
). For arbitrary set of operators $\{V_{\alpha }\},$ neither Eq. (\ref{M1})
nor Eq (\ref{M2}) fulfill the conditions (\ref{ConditionsClosed}).

When the operators $\{V_{\alpha }\}$ lead to a renewal measurement process
[Eq. (\ref{VRenewal})], Eq. (\ref{M1})\ becomes%
\begin{equation}
\mathbb{M}[\rho ]=\bar{\rho}_{s}\otimes \frac{\sum\nolimits_{i}^{^{\prime
}}\left\langle ua_{i}\right\vert \rho \left\vert ua_{i}\right\rangle \Pi _{i}%
}{\sum\nolimits_{i}^{^{\prime }}\left\langle ua_{i}\right\vert \rho
\left\vert ua_{i}\right\rangle }.
\end{equation}%
This structure does not satisfy the condition (\ref{ConditionsClosed}).
Nevertheless, when the ancilla Hilbert space is bidimensional, $i=0,1,$
given that $\sum\nolimits_{i}^{^{\prime }}$ excludes the contribution $i=0,$
it follows%
\begin{equation}
\mathbb{M}[\rho ]=\bar{\rho}_{s}\otimes \Pi _{1},  \label{PiUno}
\end{equation}%
which evidently satisfies Eq. (\ref{ConditionsClosed}). $\bar{\rho}_{s}$ is
given by Eq. (\ref{reseting}). Therefore, only when the kernel is an
exponential one, Eq. (\ref{kernelExpor}), a closed system (renewal)
measurement dynamics can be associated with the non-Markovian evolution (\ref%
{NoMarkovMasterUno}).

In the second case, Eq. (\ref{M2}) with the operators (\ref{VRenewal})
becomes 
\begin{subequations}
\begin{eqnarray}
\mathbb{M}[\rho ] &=&\mathcal{M}[\left\langle a_{\mathrm{0}}\right\vert \rho
_{sa}\left\vert a_{\mathrm{0}}\right\rangle ]\otimes \Pi _{0}, \\
&=&\bar{\rho}_{s}\otimes \Pi _{0}.
\end{eqnarray}%
Independently of the ancilla dimension the closure conditions (\ref%
{ConditionsClosed}) are satisfied. Thus, under the renewal condition the
non-Markovian evolution (\ref{MasterSimpler}) can be unravelled
independently of the ancilla dynamics, that is, for arbitrary kernels with
the structure defined by Eq. (\ref{LindbladKernel2}).


\subsection{Conditional dynamics}

In the bipartite description, the non-Markovian conditional system dynamic
can be obtained by tracing out the joint system-ancilla dynamics. Therefore,
it is possible to obtain alternative expressions for the system conditional
propagators written in terms of the bipartite dynamics.

In the first case, the conditional propagator $\mathcal{\hat{T}}(u),$ Eq. (%
\ref{TLapla}), from Eqs. (\ref{LindbladBipartito}) and (\ref%
{LindbladInteraction1}) can also be written as 
\end{subequations}
\begin{equation}
\mathcal{\hat{T}}(u)[\rho ]=\mathrm{Tr}_{a}\Big{[}\frac{1}{u-(\mathcal{L}%
_{s}+\mathcal{L}_{a}+\mathbb{D)}}(\rho \otimes \Pi _{\mathrm{0}})\Big{]},
\label{Tbipartitol}
\end{equation}%
where the bipartite initial condition (\ref{CISeparable}) was taken into
account. The propagator $\mathcal{\hat{T}}^{\prime }(u),$ Eq. (\ref%
{TprimaLapla}), taking into account the resetting state (\ref{PiUno}) becomes%
\begin{equation}
\mathcal{\hat{T}}^{\prime }(u)[\rho ]=\mathrm{Tr}_{a}\Big{[}\frac{1}{u-(%
\mathcal{L}_{s}+\mathcal{L}_{a}+\mathbb{D)}}(\rho \otimes \Pi _{\mathrm{1}})%
\Big{]}.
\end{equation}%
Therefore, the difference between both propagators arises from a different
ancilla initial condition. In the previous two equations, the bipartite
superoperator $\mathbb{D}$\ is defined by the expression $\mathcal{C}_{sa}=%
\mathbb{D}+\mathbb{J},$ where $\mathcal{C}_{sa}$ is given by Eq. (\ref%
{LindbladInteraction1}) and $\mathbb{J}$ defines the bipartite measurement
transformation $\mathbb{M}[\rho ]=\mathbb{J}[\rho ]/\mathrm{Tr}_{sa}[\mathbb{%
J}\rho ],$ Eq. (\ref{M1}). Therefore, it reads 
\begin{subequations}
\begin{eqnarray}
\mathbb{D}[\rho ] &=&-\frac{1}{2}\sum\nolimits_{i,\alpha }^{^{\prime
}}\gamma _{\alpha }\{T_{\alpha i}^{\dag }T_{\alpha i},\rho \}_{+}, \\
&=&-\frac{1}{2}\sum\nolimits_{\alpha }\gamma _{\alpha }\{V_{\alpha }^{\dag
}V_{\alpha }\otimes \Pi _{\mathrm{1}},\rho \}_{+}.
\end{eqnarray}%
In the second line, as well as in the previous two equations for $\mathcal{%
\hat{T}}(u)$ and $\mathcal{\hat{T}}^{\prime }(u),$ we used that the
configurational bath space is two-dimensional.

In the second case, both propagators are the same, $\mathcal{\hat{T}}%
^{\prime }(u)=\mathcal{\hat{T}}(u).$ From Eqs. (\ref{LindbladBipartito}) and
(\ref{LindbladInteraction2}), $\mathcal{\hat{T}}(u)$ can be written as in
Eq. (\ref{Tbipartitol}) with the superoperator $\mathbb{D}$ defined by the
expression 
\end{subequations}
\begin{subequations}
\begin{eqnarray}
\mathbb{D}[\rho ] &\mathbb{=}&-\frac{1}{2}\sum\nolimits_{\alpha }\gamma
_{\alpha }\{T_{\alpha 0}^{\dag }T_{\alpha 0},\rho \}_{+}, \\
&=&-\frac{1}{2}\sum\nolimits_{\alpha }\gamma _{\alpha }\{V_{\alpha }^{\dag
}V_{\alpha }\otimes \Pi _{\mathrm{0}},\rho \}_{+}.
\end{eqnarray}%
In this case, this definition is valid for an arbitrary number of
bath-states.

The previous expressions for the conditional propagators can be solved in an
exact way in the Laplace domain. They also provide an alternative and
equivalent way for getting statistical objects such as the survival
probabilities Eq. (\ref{Survivilinas}), or equivalently their associated
waiting time densities, Eqs. (\ref{waiter}) and (\ref{waiterIn}).

\end{subequations}

\end{document}